\documentclass{aa}  

\usepackage{graphicx}
\usepackage{txfonts}
\usepackage{xcolor}
\begin{document}

   \title{Multi-waveband detection of quasi-periodic pulsations in a stellar flare on EK Draconis observed by XMM-\emph{Newton}}
\titlerunning{QPPs in a flare on EK Dra}
\authorrunning{Broomhall et al.}

    \author{A.-M. Broomhall\inst{1, 2, 3}
         \and
			A.E.L. Thomas\inst{4, 1} \and
         C.E. Pugh\inst{1,3}
         \and
         J.P. Pye\inst{5}\and
         S.R. Rosen\inst{5,6}
          }

   \institute{Department of Physics, University of Warwick, Coventry, CV4 7AL, UK
            \and
            Institute of Advanced Study, University of Warwick, Coventry, CV4 7HS, UK
             \and
            Centre for Habitability, University of Warwick, Coventry, CV4 7HS, UK
			\email{a-m.broomhall@warwick.ac.uk}
			\and
			School of Physics and Astronomy, University of Birmingham, Edgbaston, Birmingham, B15 2TT, UK
            \and
            University of Leicester, Department of Physics \& Astronomy, Leicester, LE1 7RH, UK
			\and
			XMM-Newton Science Operations Centre, ESA, Villafranca del Castillo, Apartado 78, E-28691 Villanueva de la Ca\~{n}ada, Spain}

   \date{Received February 30, 2017; accepted April 1, 2017}

 
  \abstract
   {Quasi-periodic pulsations (QPPs) are time variations in the energy emission during a flare that are observed on both the Sun and other stars and thus have the potential to link the physics of solar and stellar flares.}
   {To characterise the QPPs detected in an X-ray flare on the solar analogue, EK Draconis, which was observed by XMM-Newton. }
   {We use wavelet and autocorrelation techniques to identify the QPPs in a detrended version of the flare. We also fit a model to the flare based on an exponential decay combined with a decaying sinusoid. The flare is examined in multiple energy bands.}
   {A statistically significant QPP is observed in the X-ray energy band of 0.2-12.0\,keV with a periodicity of $76\pm2\,\rm min$. When this energy band is split, a statistically significant QPP is observed in the low-energy band (0.2-1.0\,keV) with a periodicity of $73\pm2\,\rm min$ and in the high-energy band (1.0-12.0\,keV) with a periodicity of $82\pm2\,\rm min$. When fitting a model to the time series the phases of the signals are also found to be significantly different in the two energy bands (with a difference of $1.8\pm0.2\,\rm rad$) and the high-energy band is found to lead the low-energy band. Furthermore, the first peak in the cross-correlation between the detrended residuals of the low- and high-energy bands is offset from zero by more than $3\sigma$ $(4.1\pm1.3\,\rm min)$. Both energy bands produce statistically significant regions in the wavelet spectrum, whose periods are consistent with those listed above. However, the peaks are broad in both the wavelet and global power spectra, with the wavelet showing evidence for a drift in period with time, and the difference in period obtained is not significant. The offset in the first peak in the cross-correlation of the detrended residuals of two non-congruent energy bands ($0.5-1.0\,\rm keV$ and $4.5-12.0\,\rm keV$) is found to be even larger ($10\pm2\,\rm min$). However, the signal-to-noise in the higher of these two energy-bands, covering the range $4.5-12.0\,\rm keV$, is low.}
   {The presence of QPPs similar to those observed on the Sun, and other stars, suggests that the physics of flares on this young solar analogue is similar to the physics of solar flares. It is possible that the differences in the QPPs detected in the two energy bands are seen because each band observes a different plasma structure. However, the phase difference, which differs more significantly between the two energy bands than the period, could also be explained in terms of the Neupert effect. This suggests that QPPs are caused by the modulation of the propagation speeds of charged particles.}

               
   \maketitle
%

\section{Introduction}
Flares have been observed on a diverse range of stars for many years. However, interest in stellar flares has recently
increased because of observations on solar-like stars of flares far larger than even the largest Earth-directed flare
ever observed on our own Sun \citep{2012Natur.485..478M}. This has raised the debate over whether a solar superflare could occur. Based on \textit{Kepler} observations, \citet{2015EP&S...67...59M} estimate that a superflare could materialise on the Sun once every 500-600yr, but to accumulate enough energy would require a sunspot to exist for a number of years. The
number of superflares observed on solar-like stars remains low and it is still uncertain how analogous these flares are to
those observed on the Sun. Furthermore, since the energies associated with superflares are orders of magnitude
greater than typical solar flares, questions over the validity of such predictions remain \citep[e.g.][]{2015JPhCS.632a2058H}. It is, therefore, necessary to ascertain the relationship between solar and stellar flares in order to truly exploit the solar-stellar connection.

Coronal seismology has the potential to provide important insights as to the link between the physical processes
observed in solar and stellar flares and their associated active regions. Solar coronal seismology, which studies
waves and oscillations in coronal plasma, is a relatively novel research field that has blossomed with the advent of
new high-quality data. The natural extension to stellar coronal seismology has yet to be fully explored. A promising route towards linking the physics of solar and stellar flares, within the remit of coronal seismology, is through the study of quasi-periodic pulsations (QPPs), which are time variations in the energy emission during a flare. There is increasing evidence to support the theory that QPPs are a common feature of both solar and stellar flares \citep{2010SoPh..267..329K, 2013ApJ...777..152S, 2016ApJ...833..284I, 2016MNRAS.459.3659P, 2017A&A...608A.101P}. For example, they were recently observed in the largest solar flare of cycle 24 to date \citep{2018ApJ...858L...3K}. This X9.3 class flare was estimated to have an energy of around $10^{32}$\,erg, and therefore bridges the energy gap observed between solar and stellar flares. 

Although observations of QPPs in stellar flares are becoming more abundant \citep[e.g.][]{2016MNRAS.459.3659P, 2018MNRAS.475.2842D}, the majority of these flares occurred on M dwarfs, which tend to be more magnetically active than the Sun. However, magnetic activity is also a function of age: In general, younger stars rotate faster than the Sun and, so, their dynamos are able to produce far stronger magnetic fields \citep{2014MNRAS.441.2361V}. As a star ages, angular momentum is lost through a magnetised stellar wind, causing the rotation rate to slow and magnetic activity to decrease. Such high levels of activity on our young Sun could have had important consequences for the early evolution of our Solar System, and similarly activity on other young Suns may affect the habitability of exoplanets. It is therefore important to understand the evolution of the magnetic fields in solar-like stars and, particularly, whether the flaring events on young stellar analogues are governed by the same physical processes as the flares observed on our Sun today.

Here we detect, in X-rays, QPPs in a flare that occurred on EK Draconis (EK Dra, HD 129333), which is considered to be a young solar analogue (G1.5V) star. Gaia DR2 gives $T_{\textrm{eff}}=5583_{-193}^{+155}\,\rm K$, $L=0.8989\pm0.0017L_{\textrm{Sun}}$, and $R=1.01_{-0.04}^{+0.07}R_{\textrm{Sun}}$ \citep[e.g][]{2016A&A...595A...1G, 2018A&A...616A...1G, 2018A&A...616A...4E}. EK Dra is thought to be part of the Pleides moving group, which is believed to have an age of $\sim100\,\rm Myr$ \citep{2005ApJ...622..680R}. Its classification as a young solar analogue means that EK Dra has been extensively studied \citep[e.g.][]{1987AJ.....93..920S, 1990AJ....100..818E, 1994ApJ...428..805D, 1995A&A...301..201G, 1999ApJ...513L..53A, 2003ApJ...594..561G, 2005A&A...432..671S, 2010ApJ...723L..38A, 2012ApJ...745...25L, 2016A&A...593A..35R, 2017A&A...599A.127F}. Recently, \citet{2017MNRAS.465.2076W} used magnetic features to find an average equatorial rotation period of $2.51\pm0.08\,\rm d$, with evidence of strong solar-like differential rotation. Since EK Dra is younger and faster rotating than the Sun it is expected to be far more magnetically active. Indeed, in two observations, \citet{2016A&A...593A..35R} found the mean field strength of EK Dra to be 66\,G and 89\,G respectively (in comparison to a mean magnetic field strength of $\sim1\,\rm G$ typically observed for the Sun). 

We introduce an intriguing aspect of these QPP detections by splitting the data into two X-ray photon-energy bands: the phase and period of the QPPs are found to be significantly different in two congruent but independent energy bands. This may be evidence for the Neupert effect \citep{1968ApJ...153L..59N}, the empirical relationship between soft X-rays and the cumulative time integral of hard X-rays during a flare, which demonstrates the direct causal relationship between energetic electrons and thermal plasma emissions. Alternatively the different periods and phases could be observed because the physical processes responsible for the QPPs are occurring in more than one plasma structure. The structure of the paper is as follows: first, in Sect. \ref{section[data]}, we describe the data used in this study. We then outline the data analysis procedures employed (Sect. \ref{section[analysis]}) and the main results (Sect. \ref{section[results]}). Finally, we discuss the implications and interpretation of these results and compare with previous works (Sect. \ref{section[discussion]}), before giving a brief summary in Sect. \ref{sec:summary}.  

\section{Data} \label{section[data]}
A flare was observed on EK Dra by XMM-Newton on December $30^{\rm th}$ 2000 14:38:24UTC\footnote{Based on observations obtained with XMM-Newton, an ESA science mission with instruments and contributions directly funded by ESA Member States and NASA} \citep[2XMM DETID 175724 and XMM observation ID 0111530101;][]{2005A&A...432..671S, 2005ApJ...622..653T, 2008A&A...482..639N, 2015A&A...581A..28P}, with an energy of $3.687\times10^{33}\textrm{erg}$. The flare was observed by the XMM-Newton European Photon Imaging Camera (EPIC), which consists of 2 Metal Oxide Semi-conductor (MOS) CCD arrays \citep{2001A&A...365L..27T} and EPIC-pn, which contains 12 pn-CCDs \citep{2001A&A...365L..18S}. All operate in photon counting mode and so register position, arrival time and energy of incoming photons, and the CCDs are sensitive to the 0.2-12.0keV range. The count rate of EPIC-pn is higher than the MOS instruments. Although this study will focus on the EPIC-pn data, we did analyse the MOS data and found consistent results. Initially we considered data in the `Total' energy band (which covers 0.2-12.0\,keV). The cadence of time series analysed here was $10\,\rm s$. Although we note that EPIC-pn is capable of shorter cadences, a cadence of 10\,s ensures a good count rate and, therefore, signal to noise, while still being sufficiently short to study the QPPs. We then split the data into smaller energy bands. Initially the data were split into two congruent energy bands: the `low'-energy band (0.2-1.0\,keV) and the `high'-energy band (1.0-12.0\,keV). Notice that in terms of energy range the higher energy band is far wider, however, since the number of photons is lower in the higher energy band this ensures  a good number of photons in each band. Again the cadence of these time series was $10\,\rm s$. Next the data were split into non-congruent energy bands, specifically 0.5-1.0\,keV and 4.5-12.0\,keV. Here a longer integration time was required to produce sufficient numbers of photons and so the cadence of these time series was $20\,\rm s$.

\section{Data analysis}\label{section[analysis]}
\begin{figure*}
   \centering
    \includegraphics[width=0.45\textwidth]{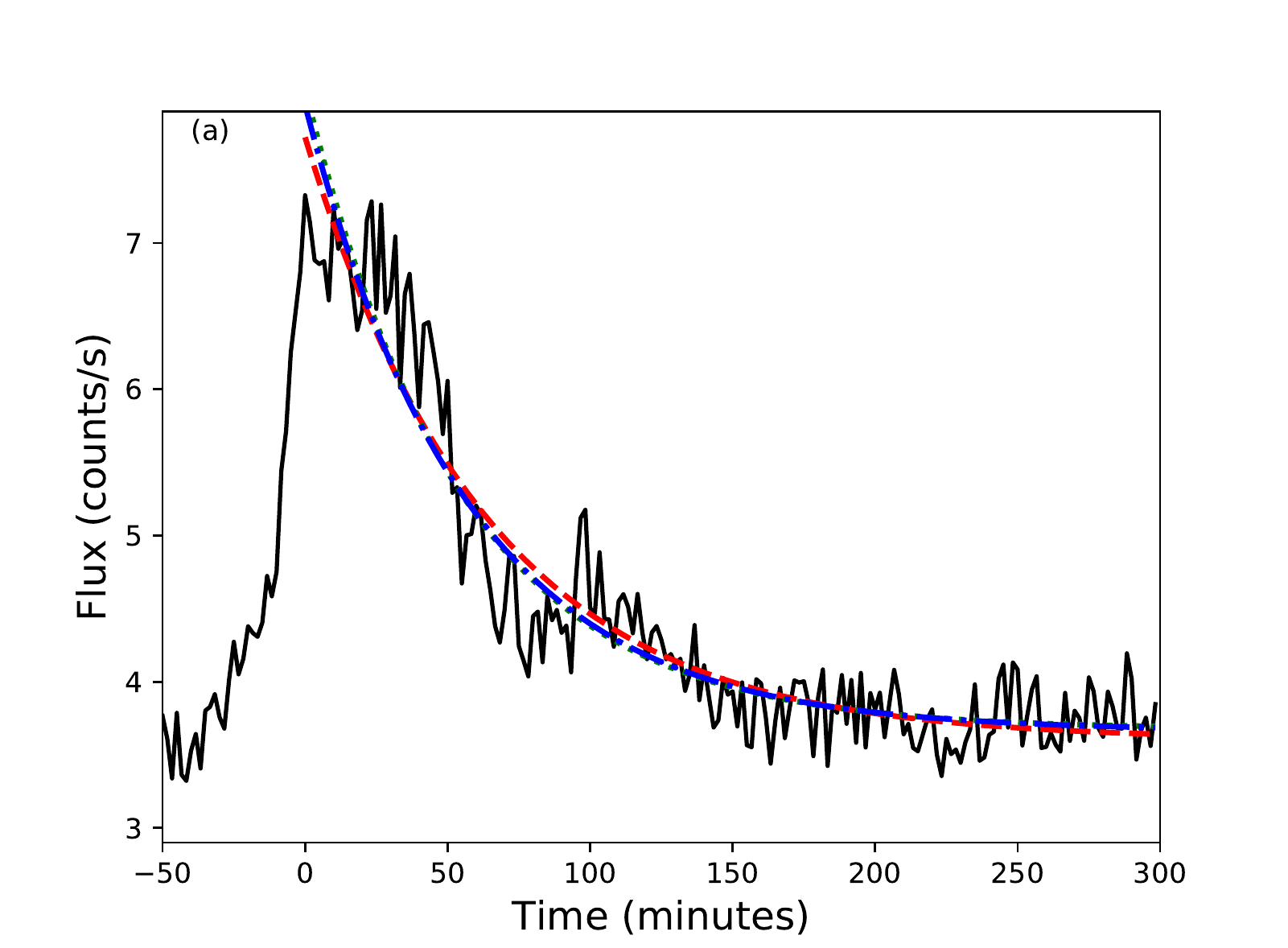}
    \includegraphics[width=0.45\textwidth]{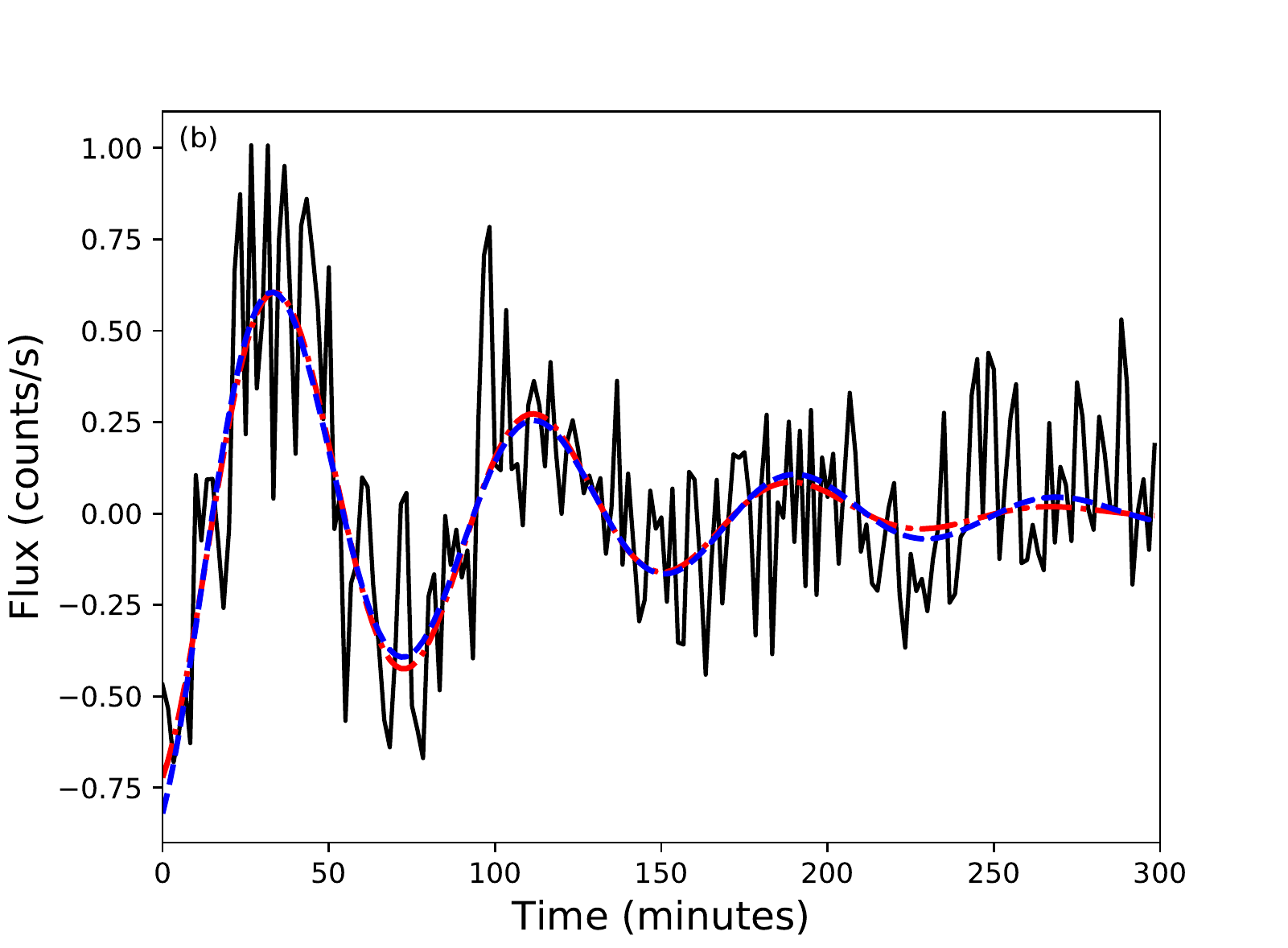}\\
    \includegraphics[width=0.45\textwidth]{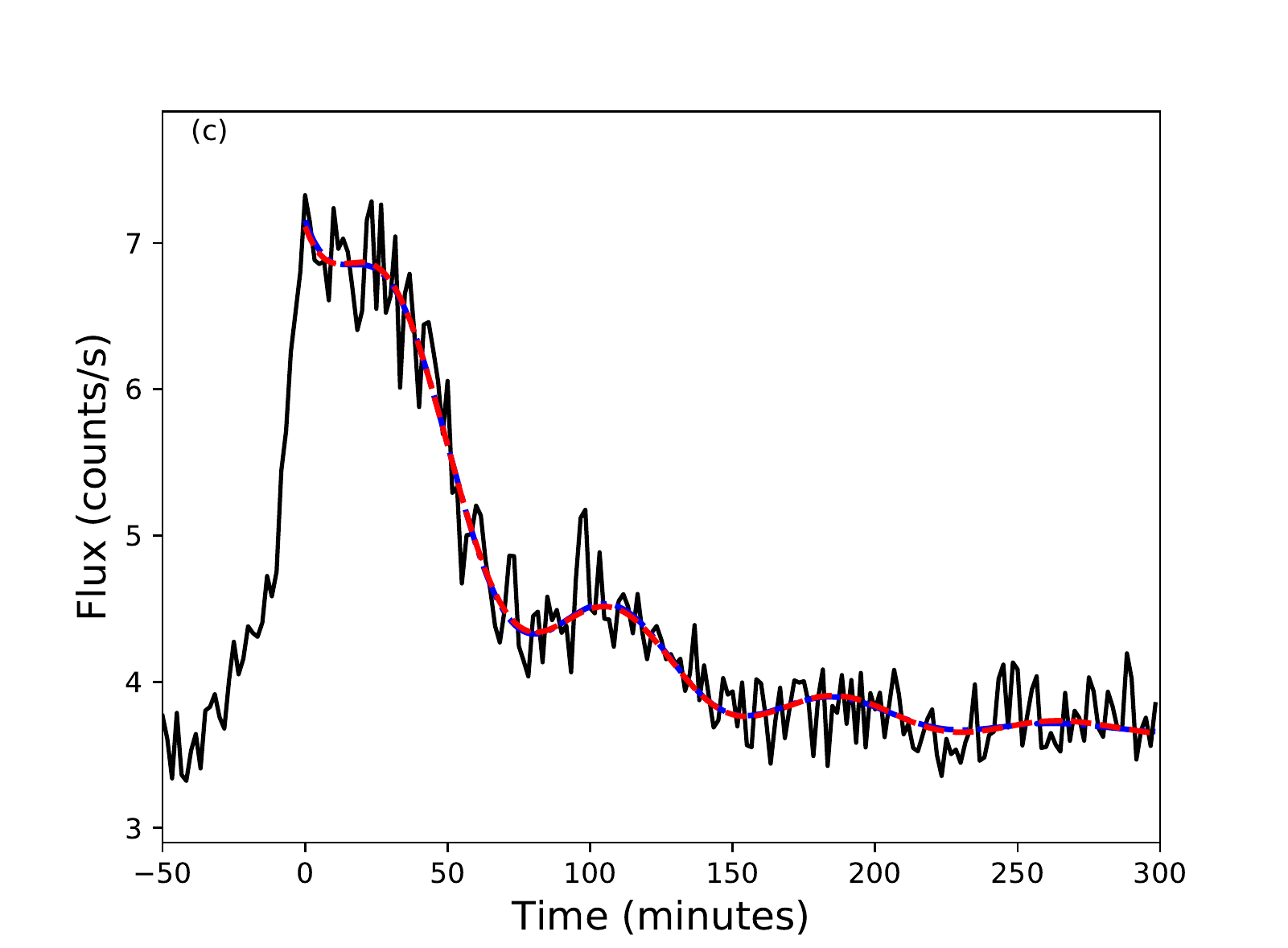}
    \includegraphics[width=0.45\textwidth]{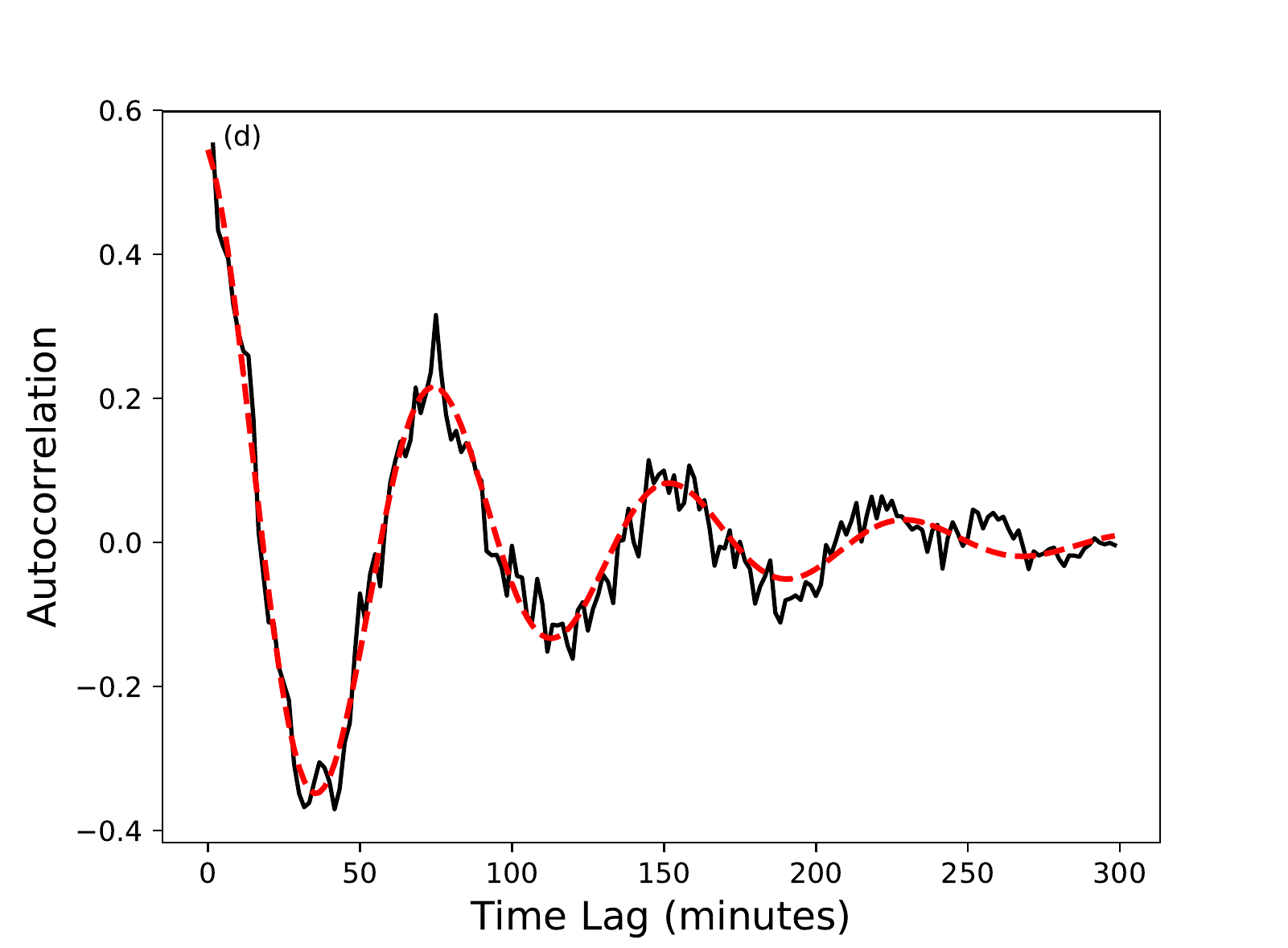}\\
   \caption{Panel (a): Flare lightcurve (black, solid) and fit to exponential decay (red, dashed). The green, dotted curve shows the exponential-decay component of the combined fit of exponentially decaying QPP and exponential decay (`full-flare exponential' fit), while the blue dot-dashed curve shows the exponential-decay component of the combined fit of Gaussian-decaying QPP and exponential decay (`full-flare Gaussian' fit). The data are from the entire 0.2-12keV range. The blue and green curves are almost identical and, therefore, difficult to distinguish. Panel (b): Detrended flare lightcurve  (black, solid). Detrending was performed by subtracting the exponential decay fit plotted in Panel (a) from flare lightcurve also plotted in Panel (a). The red, dashed curve shows an exponential decaying sinusoid, while the blue, dot-dashed curve shows a Gaussian decay in the sinusoid, both of which were fitted to the detrended flare lightcurve. The blue and red curves are almost identical and, therefore, difficult to distinguish. Panel (c): Flare lightcurve (black, solid) and fits to lightcurve consisting of the sum of exponential decay term and decaying sinusoid. The red, dashed line shows the fit when the decay of the sinusoid was described by an exponential, while the blue, dot-dashed curved shows the fit when the decay of the sinusoid was described by a Gaussian. The blue and red curves are almost identical and, therefore, difficult to distinguish. Panel (d): Autocorrelation of detrended lightcurve (black solid line) from panel (b). An exponentially decaying lightcurve was fitted to the autocorrelation (red, dashed line).}
              \label{figure[flare]}%
\end{figure*}

The methodology used to study this flare is based upon that utilised by \citet{2015ApJ...813L...5P} and \citet{2016MNRAS.459.3659P} to study QPPs in flares observed by NASA's \textit{Kepler} spacecraft, however, we now outline the procedure for clarity. In order to deduce whether periodic behaviour is present in the light curve it was first necessary to remove the decay trend of the flare from each data set. This was done by fitting the following expression to the decaying phase of the flare using a least-squares method:

\begin{equation}
F(t) = A_0\exp\left(-\frac{t}{t_{0}}\right)+C,
\label{eqn: decay}
\end{equation}

\noindent where $F$ is the flux, $t$ is the time, $t_0$ is the e-folding time, $A_0$ is the amplitude of the flare, and $C$ is a constant, which characterises the quiescent flux. The decay phase of the flare and the fitted curve are shown in Fig. \ref{figure[flare]}a. Uncertainties on the fitted parameters were obtained through 5,000 Monte Carlo simulations where the observed flux values were modified by Gaussian noise and the width of the Gaussian distribution for each data point was determined by the formal uncertainties associated with the data. For each realisation, a fit to the data was performed. Once all 5,000 realisations had been fitted, histograms of the output parameters were produced and examples can be seen in Appendix \ref{section[histograms]}. We note here that the same 5,000 Monte Carlo simulations were used to determine uncertainties on fitted parameters in all subsequent analysis steps as described below, including the cross-correlation analysis. For the majority of parameters Gaussian curves were fitted to the histograms and the widths of the Gaussians were then used to determine the uncertainties on the fitted parameters. Although when fitting Eq. \ref{eqn: decay} to the flux rates Gaussian curves were good fits to the histograms, this was not always the case when determining QPP parameters (see Sect. \ref{section[results_total]} and Appendix \ref{section[histograms]}). Therefore, we also give the median and quartile values of the distribution of the Monte Carlo simulation results for each parameter.  
 
Once a fit was found, the decay given by Eq. \ref{eqn: decay} was subtracted from the flare flux rate to produce a detrended light curve, as seen in Fig. \ref{figure[flare]}b. For the remainder of this paper we refer to detrended light curves as residuals. The residuals appear to exhibit a QPP-like signal. We note that although we have removed the dominant background trend caused by the flare itself, there could still be some form red noise signal in the data. We therefore ascribe the observed quasi-periodic signal to QPPs but we note that there could be numerous explanations for this QPP. Potential QPP excitation mechanisms are discussed in \citep{2018SSRv..214...45M}, although even for the Sun, where it is possible to make resolved observations of the active region, it is still difficult to differentiate between these excitation mechanisms. As we do not have resolved observations here there is no way of unequivocally linking the QPPs to the active region responsible for the flare. However, the fact that the QPPs coincide with the flare and are not observed in the time series away from the flare is a good indication that the two are related.

The QPP signal can be characterised by a decaying sinusoid. It has recently been demonstrated for solar data that magnetohydrodynamic oscillations can be characterised by both exponential and Gaussian decay. An exponential decay is the more traditional assumption and was used by \citet{2016ApJ...830..110C} to fit QPPs observed in XMM-Newton data, where the flare had been detrended using Empirical Mode Decomposition. However, Gaussian decay or a combination of Gaussian and exponential decay phases has recently been justified for magnetohydrodynamic oscillations through numerical simulations and analytically \citep{2012A&A...539A..37P, 2013A&A...551A..39H, 2016A&A...585L...6P, 2017A&A...600A..78P}. Since at this point we do not know the origin of the observed QPPs, we attempted to fit both cases to the residuals, using least-squares methods. The exponentially decaying sinusoid fitted to the residuals is given by:
\begin{equation}
R(t) = A_e\exp\left(-\frac{(t-B_e)}{\tau_e}\right)\cos\left(\frac{2\pi t}{P_e}+\phi_e\right),
\label{eqn: oscillations exp}
\end{equation}
where $R(t)$ is the residual, $A_e$ is the amplitude, $t$ is the time, $B_e$ is a constant, $\tau_e$ is the exponential damping time, $P_e$ is the period, and $\phi_e$ is the phase. The Gaussian decaying sinusoid fitted to the residuals is given by:
\begin{equation}
R(t) = A_g\exp\left(-\frac{(t-B_g)^2}{2\tau_g^2}\right)\cos\left(\frac{2\pi t}{P_g}+\phi_g\right),
\label{eqn: oscillations gauss}
\end{equation}
where $A_g$ is the amplitude, $\tau_g$ is the Gaussian damping time, $P_g$ is the period, $\phi_g$ is the phase, and $B_g$ is a constant allowing the peak of the Gaussian envelope to be offset from $t=0$. Uncertainties on the fitted parameters were determined using Monte Carlo simulations, as described above. 

Using the parameters obtained from fitting Eqs. \ref{eqn: decay}, \ref{eqn: oscillations exp}, and \ref{eqn: oscillations gauss} as initial guesses, the original data from the flare decay were fitted with a summation of the decay phase described by Eq. \ref{eqn: decay} and a sinusoidal component given by either Eq. \ref{eqn: oscillations exp} or Eq. \ref{eqn: oscillations gauss}. For the remainder of the paper these fits will be referred to as `full-flare exponential' or `full-flare Gaussian' fits respectively. The fitting was conducted using a least-squares method and as before the uncertainties on the fitted parameters were estimated using the Monte Carlo simulations outlined above. Both fits can be seen in Fig. \ref{figure[flare]}c.

The time-lagged autocorrelation of data is a useful tool in highlighting oscillatory behaviour in a signal as noise tends to be suppressed. Figure \ref{figure[flare]}d shows the autocorrelation of the residuals plotted in Fig. \ref{figure[flare]}b. Again a decaying oscillatory signal is observed and so an exponentially decaying sinusoid was also fitted to the autocorrelation using the following expression:

\begin{equation}
\mathcal{C}(t) = A_a\exp\left(-\frac{t-B_a}{\tau_a}\right)\cos\left(\frac{2\pi t}{P_a}+\phi_a\right),
\label{eqn: auto}
\end{equation} 

\noindent where $\mathcal{C}(t)$ is the correlation, $t$ is the time lag from zero where the signal is perfectly correlated with itself, $A_a$ is the amplitude of the signal, $\tau_a$ is the exponential decay time, $P_a$ is the period, $\phi_a$ is the phase, and $B_a$ is a constant. As described above, Monte Carlo simulations were used to determine the parameter values, uncertainties and stability of the fits.

To determine the significance of the oscillatory signal in the residuals and to provide an alternative measure of its period a wavelet transform was performed \citep{Torrence98apractical}. For this study the Morlet wavelet was used since it gave plots with the best balance between time and period resolutions. With a wavelet transform it is also possible to assess how stable the period of a signal is in time, something that was assumed when fitting Eqs. \ref{eqn: oscillations exp} and \ref{eqn: oscillations gauss} to the data. The time series was padded with zeroes at the start to double the length of observation. As the QPP signal has a maximum amplitude at the peak of the flare, that is at the start of the time series we consider, this shifts the signal away from the cone of influence. A signal was considered significant if it was above the 99\% confidence level. These significance levels were determined based on an assumption of white noise and a confidence level of 99\% indicates that there is less than a 1\% chance of observing a signal of this amplitude if the data only contain white noise. This was the approach taken in \citet{2016MNRAS.459.3659P}, upon which this work is based, but is also adopted in other QPP studies \citep[e.g.][]{2010SoPh..267..329K, 2011ApJ...740...90V, 2013ApJ...773..156A}. In addition to the standard confidence limits, we also determined significance levels using the recommendations of \citet{2016ApJ...825..110A} which are also based on an assumption of white noise but take into account the total number of degrees of freedom of the wavelet spectra. 

As a double check on the significance levels we utilised a Fisher Randomisation Test \citep{fisher:1935}. The randomisation was performed 5,000 times and used to determine the significance of features in the wavelet spectrum (see Appendix \ref{sec: extra_wavelets} for details). Although we only show the wavelet for the total energy band, these randomisation tests confirm the significance of peaks observed in the other energy bands as well. As it is possible that some red noise remains in the residuals we also tested the significance of the signal based on an assumption of red noise (also described in Appendix \ref{sec: extra_wavelets}). In the remainder of this article all signals found to be significant based on the white noise assumption were also found to be significant based on the red noise hypothesis. We also note here that significant peaks were found at consistent periods in the MOS data and while it is possible that some of the noise may be coherent between the MOS and EPIC-pn data, particularly that of stellar origin, this adds weight to the detection. 

In order to deduce a mean period from the wavelet transform, a global wavelet spectrum was plotted which takes a time-average of the transform. Again we determine both the traditional white noise 99\% significance levels and those recommended by \citet{2016ApJ...825..110A}. A Gaussian fit was applied to the main peak in order to estimate the peak period (see right-hand panel of Fig. \ref{figure[wavelet]}) and the width of this peak was used to determine the uncertainty on the period.

\section{Results} \label{section[results]}
\subsection{Total energy band}\label{section[results_total]}

Panel (a) of Fig. \ref{figure[flare]} shows the exponential fit to the decay phase of the flare and the fitted parameters are given in Table \ref{table[flare_decay_fits]}. The histograms produced by the Monte Carlo simulations are well-represented by a Gaussian shape (see Appendix \ref{section[histograms]}), and the medians (and interquartile ranges) give very similar parameter values to those obtained by fitting a Gaussian to the histograms.

\begin{table}\caption{Parameters obtained when fitting Eq. \ref{eqn: decay} to EPIC-pn data for total energy band.}\label{table[flare_decay_fits]}
\centering
\begin{tabular}{ccc}
  \hline
Parameter & Histogram fit  & Median \\
  \hline
$A_0$ (counts/s) &  $4.12\pm0.07$ & $4.12\pm^{0.04}_{0.05}$ \\
$t_0$ (min) &  $63.8\pm1.7$ & $63.8\pm1.1$ \\
$C$ (counts/s) &  $3.60\pm0.02$ & $3.60\pm0.01$ \\
  \hline
\end{tabular}
\end{table}

The residuals, shown in panel (b) of Fig. \ref{figure[flare]}, are well fit by both the exponentially and Gaussian decaying sinusoids. However, the Monte Carlo simulations reveal that the exponentially decaying fit is far more stable: for the Gaussian-decaying sinusoid tight bounds needed to be placed on the parameter space to ensure the majority of fits converged. Nevertheless, over 1,400 of the 5,000 Monte Carlo simulation fits failed. Conversely, all 5,000 Monte Carlo simulations were successfully fitted with the exponentially-decaying sinusoid. This implies that the exponentially decaying sinusoid is a better representation of the data given observational uncertainties on each data point. Appendix \ref{section[histograms]} shows the histograms that were produced from the fitted parameters. When an exponentially-decaying sinusoid was fitted to the data, symmetric Gaussian-shaped histograms were produced for all parameters. However, when a Gaussian-decaying sinusoid was fitted to the data skewed histograms were produced for $B_g$, and also $A_g$ and $\tau_g$ when plotted in linear space. This implies that for these parameters symmetric uncertainties are inappropriate and we therefore take the median values and interquartile ranges to be the output fit values and uncertainties respectively. The fitted values of the parameters can be found in Table \ref{table[flare_qpp_fits]}.  Panel (b) of Fig. \ref{figure[flare]} shows that both the exponentially-decaying and Gaussian-decaying sinusoid fits are similar and indeed the periodicities of the sinusoids in the two fits are in very good agreement, producing values of $78.7^{+0.9}_{-0.8}$\,min for the exponentially-decaying sinusoid and $78.1^{+1.1}_{-1.2}$\,min for the Gaussian-decaying sinusoid. 

\begin{table}\caption{Comparison of parameters obtained when fitting Eqs. \ref{eqn: oscillations exp} and  \ref{eqn: oscillations gauss} to residuals for total energy band.}\label{table[flare_qpp_fits]}
\centering
\begin{tabular}{ccc}
  \hline
Parameter & Histogram fit  & Median \\
  \hline
$A_e$ (counts/s) &  $0.85\pm0.08$ & $0.85\pm0.05$ \\
$B_e$ (counts/s) &  $1.9\pm1.9$ & $1.9\pm^{+1.3}_{-1.4}$ \\
$\tau_e$ (min) &  $95\pm10$ & $95\pm7$ \\
$P_e$ (min) & $78.7\pm1.3$ & $78.8^{+0.9}_{-0.8}$ \\
$\phi_e$ (radians) & $3.54\pm0.10$ & $3.53^{+0.6}_{-0.7}$\\
\hline
$A_g$ (counts/s) &  $0.71\pm0.09$ & $0.72^{+0.8}_{-0.6}$ \\
$B_g$ (counts/s) &  $10\pm35$ & $2^{+19}_{-31}$ \\
$\tau_g$ (min) &  $79\pm25$ & $79^{+20}_{-14}$ \\
$P_g$ (min) & $78.1\pm1.7$ & $78.1\pm^{+1.1}_{-1.2}$ \\
$\phi_g$ (radians) & $3.48\pm0.11$ & $3.48\pm_{-0.08}^{+0.07}$\\
  \hline
\end{tabular}
\end{table}

Panel (c) of Fig. \ref{figure[flare]} shows the full-flare exponential and Gaussian fits. Again we can see that the fitted curves are in good agreement, as is reflected by the fitted parameters, which are given in Table \ref{table[total_fits]}. Once again the fits for the exponentially-decaying sinusoid were found to be more stable: for the Gaussian-decaying sinusoids a large number of the Monte Carlo fits failed to converge (over 2,500), compared with no failures for the exponentially-decaying sinusoid. However, this still left over 2,000 simulations to constrain the values of the parameters for the full-flare Gaussian fit. For the full-flare Gaussian fit, a number of histograms of the fitted parameters were found to be non-Gaussian or skewed Gaussian in shape (see Appendix \ref{section[histograms]}). The fitted periods were $76.1\pm1.0$\,min for the exponentially decaying sinusoid and $76.1\pm1.2$\,min for the Gaussian-decaying sinusoid, again in good agreement with each other. Although these values are systematically smaller than those found when fitting the residuals, they are still within $2\sigma$ of the residual results. Panel (a) of Fig. \ref{figure[flare]} compares the exponential-decay part of the full-flare fits with that obtained when only fitting the exponential-decay (i.e. when fitting Eq. \ref{eqn: decay} alone). All three fitted trends are in good agreement with each other. 

\begin{table}\caption{Parameters obtained when performing full-flare fits to flux for total-energy band. The parameters in the top half were found when an exponentially-decaying sinusoid was included in the full-flare fit, and the parameters in the bottom half were obtained when a Gaussian-decaying sinusoid was included in the full-flare fit. }\label{table[total_fits]}
\centering
\begin{tabular}{ccc}
  \hline
Parameter & Histogram fit  & Median \\
  \hline
$A_0$ (counts/s) & $4.37\pm0.11$ & $4.38\pm0.07$\\
$t_0$ (min) & $54.7\pm2.6$ & $54.8^{+1.8}_{-1.7}$\\
$C$ (counts/s) & $3.68\pm0.02$ & $3.68\pm0.01$\\
$A_e$ (counts/s) &  $0.748\pm0.011$ & $0.748\pm^{+0.007}_{-0.008}$ \\
$B_e$ (counts/s) &  $17.9\pm6.5$ & $17.6^{+4.1}_{-5.0}$ \\
$\tau_e$ (min) &  $88.3\pm9.4$ & $88.6^{+6.7}_{-6.6}$ \\
$P_e$ (min) & $76.1\pm1.5$ & $76.1\pm1.0$ \\
$\phi_e$ (radians) & $3.28\pm0.17$ & $3.28\pm0.12$\\
\hline
$A_0$ (counts/s) & $4.29\pm0.11$ & $4.29\pm0.07$\\
$t_0$ (min) & $56.5\pm2.0$ & $56.5\pm1.3$\\
$C$ (counts/s) & $3.67\pm0.02$ & $3.67\pm0.01$\\
$A_g$ (counts/s) &  $0.74\pm0.15$ & $0.83^{+0.32}_{-0.14}$ \\
$B_g$ (counts/s) &  $-17\pm95$ & $-43^{+46}_{-65}$ \\
$\tau_g$ (min) &  $110\pm44$ & $104^{+26}_{-25}$ \\
$P_g$ (min) & $76.1\pm1.7$ & $76.1\pm1.2$ \\
$\phi_g$ (radians) & $3.32\pm0.14$ & $3.31_{-0.11}^{+0.10}$\\
  \hline
\end{tabular}
\end{table}

Panel (d) of Fig. \ref{figure[flare]} shows the autocorrelation of the residuals. This was fitted with an exponentially decaying sinusoid and the values of the fitted parameters can be found in Table \ref{table[auto]}. The period of the sinusoid was found to be $78.0^{+0.9}_{-0.8}$\,min, in good agreement with the values obtained by fitting the residuals themselves. 

\begin{table}\caption{Parameters obtained when fitting Eq. \ref{eqn: auto} to autocorrelation of residuals. }\label{table[auto]}
\centering
\begin{tabular}{ccc}
  \hline
Parameter & Histogram fit  & Median \\
  \hline
$A_a$ (counts) & $0.221\pm0.015$ & $0.221^{+0.018}_{-0.002}$\\
$B_a$ (counts) &  $56\pm8$ & $56^{+6}_{-5}$ \\
$\tau_a$ (min) &  $66\pm8$ & $66\pm5$ \\
$P_a$ (min) & $78.0\pm1.3$ & $78.0^{+0.9}_{-0.8}$ \\
$\phi_a$ (radians) & $0.11\pm0.06$ & $0.12\pm0.04$\\
  \hline
\end{tabular}
\end{table}

\begin{figure*}
   \centering
    \includegraphics[width=0.7\textwidth]{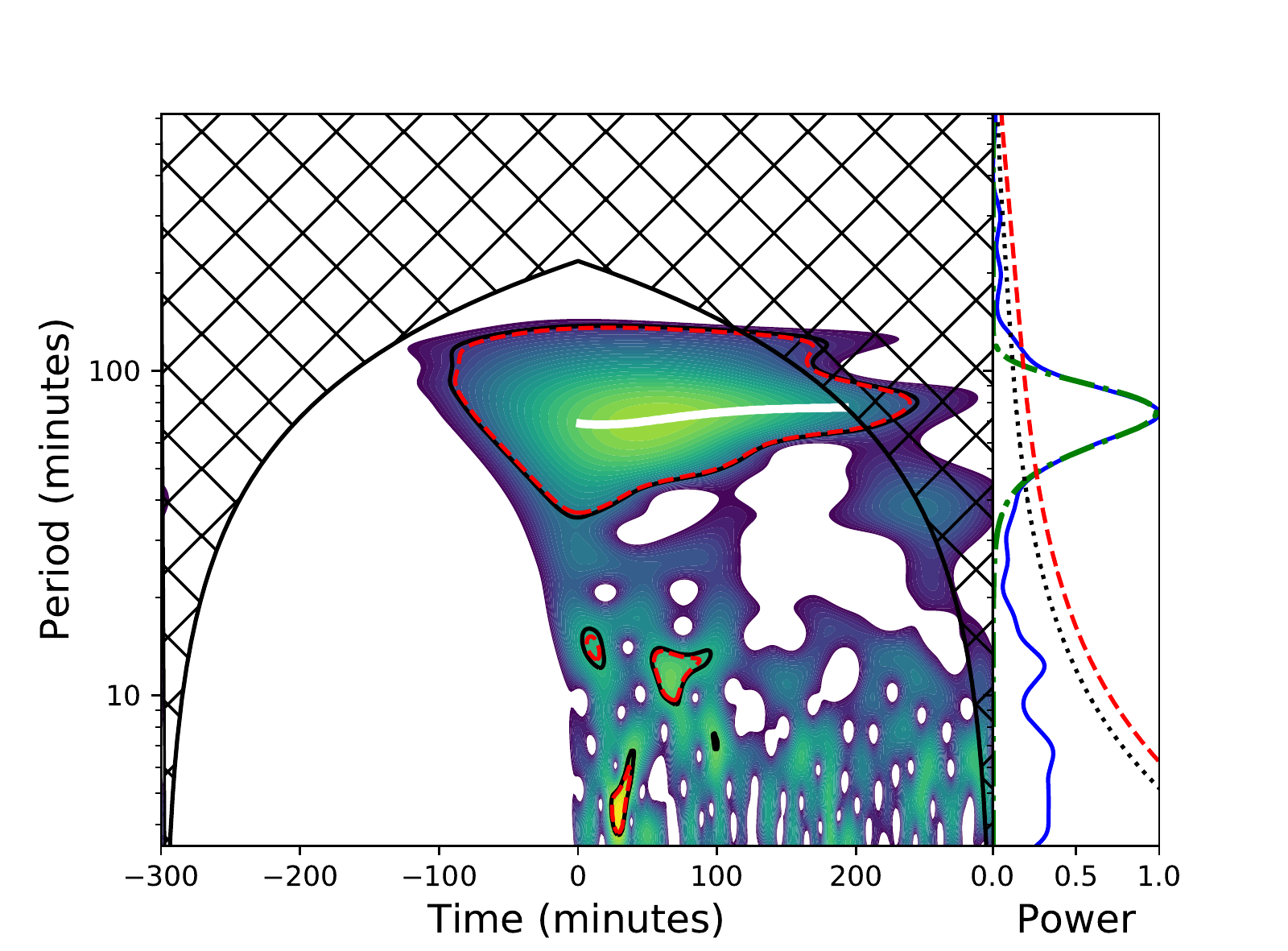}\caption{Wavelet transform of residuals for total energy band. Black solid contours indicate the standard 99\% significance levels, while red dashed contours indicate 99\% significance levels modified by the recommendations of \citet{2016ApJ...825..110A}. The white line indicates the ridge of maximum power, which evolves from $70\,\rm min$ at $t=0\,\rm min$ to $77\,\rm min$ at $t=190\,\rm min$. Black hatching and associated arcs indicate the cone of influence, where edge effects become important. Observed features confined to this cone are disregarded. Also plotted on the right is the global wavelet. The black-dotted line indicates the standard 99\% significance level and the red-dashed line indicates the modified 99\% significance level. The green dot-dashed curve represents a best fit to the main peak based on a Gaussian shape.}
              \label{figure[wavelet]}%
\end{figure*}

Figure \ref{figure[wavelet]} shows the wavelet transform of the residuals. There is a broad peak that is significant above the 99\% significance level with a periodicity of approximately 70\,mins, and that lasts for $\sim250\,\rm mins$ (and into the cone of influence). Fitting a Gaussian shape to the corresponding peak in the global power spectrum gives a period of $74\pm16$\,min, where the uncertainties are based upon the fitted width of the peak. Again this is in good agreement with the values obtained by fitting the residuals, the autocorrelation and the full flare. However, we note that the uncertainties associated with this period are far larger than those obtained from fitting the data. This is likely to reflect the fact that the period drifts with time: the ridge of maximum power in the wavelet evolves from around $70\,\rm min$ at $t=0\,\rm min$ to around $77\,\rm min$ at around $t=190\,\rm min$ (at the edge of the cone of influence).  In Eqs. \ref{eqn: oscillations exp} and \ref{eqn: oscillations gauss} we assume a stationary period. The wavelet analysis, therefore, implies that the uncertainties associated with the fitted period may have been underestimated.

\subsection{Results obtained when splitting the data into two congruent energy bands}\label{section[congruent]}
    
\begin{figure*}
   \centering
    \includegraphics[width=0.45\textwidth]{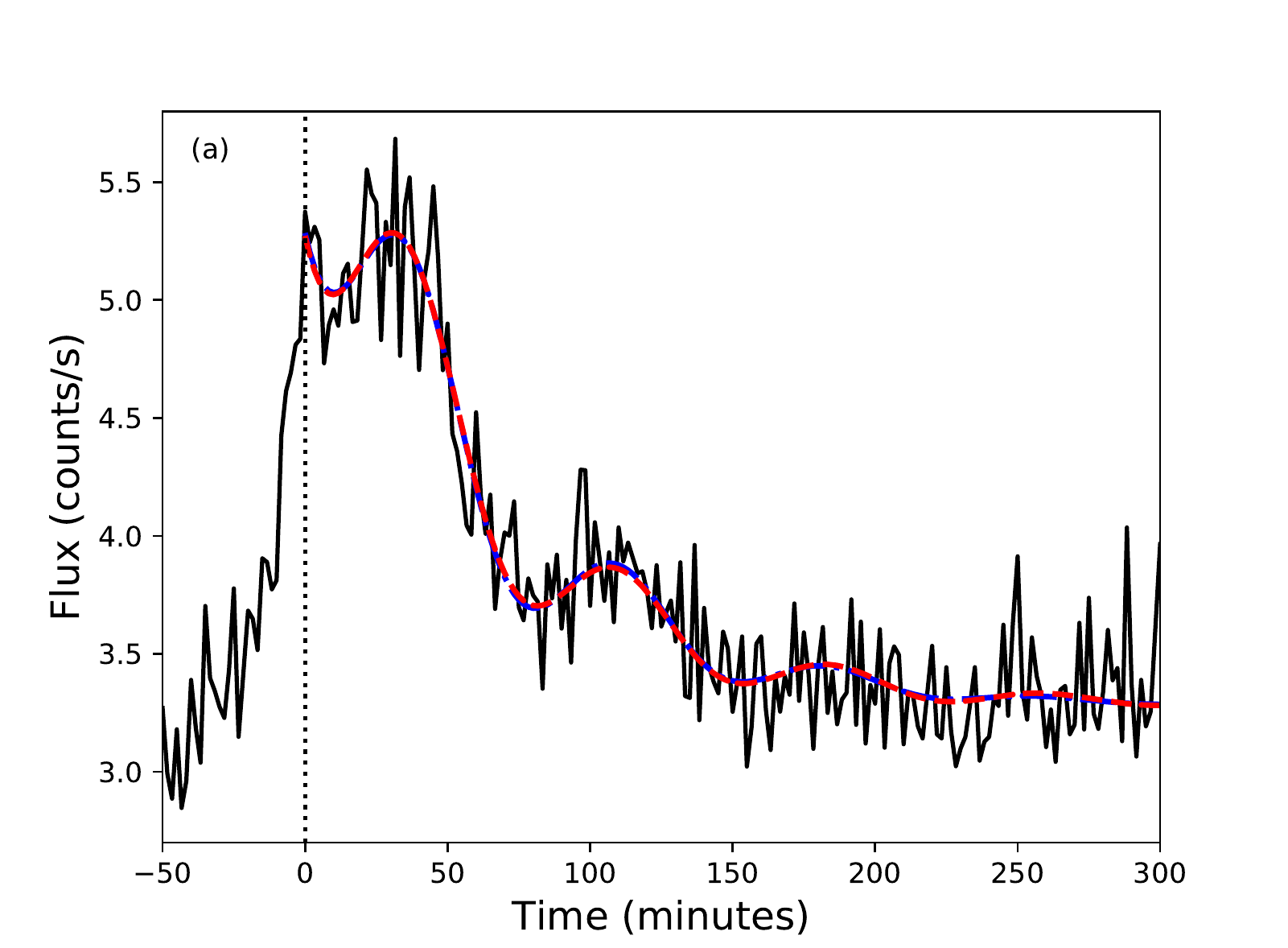}
    \includegraphics[width=0.45\textwidth]{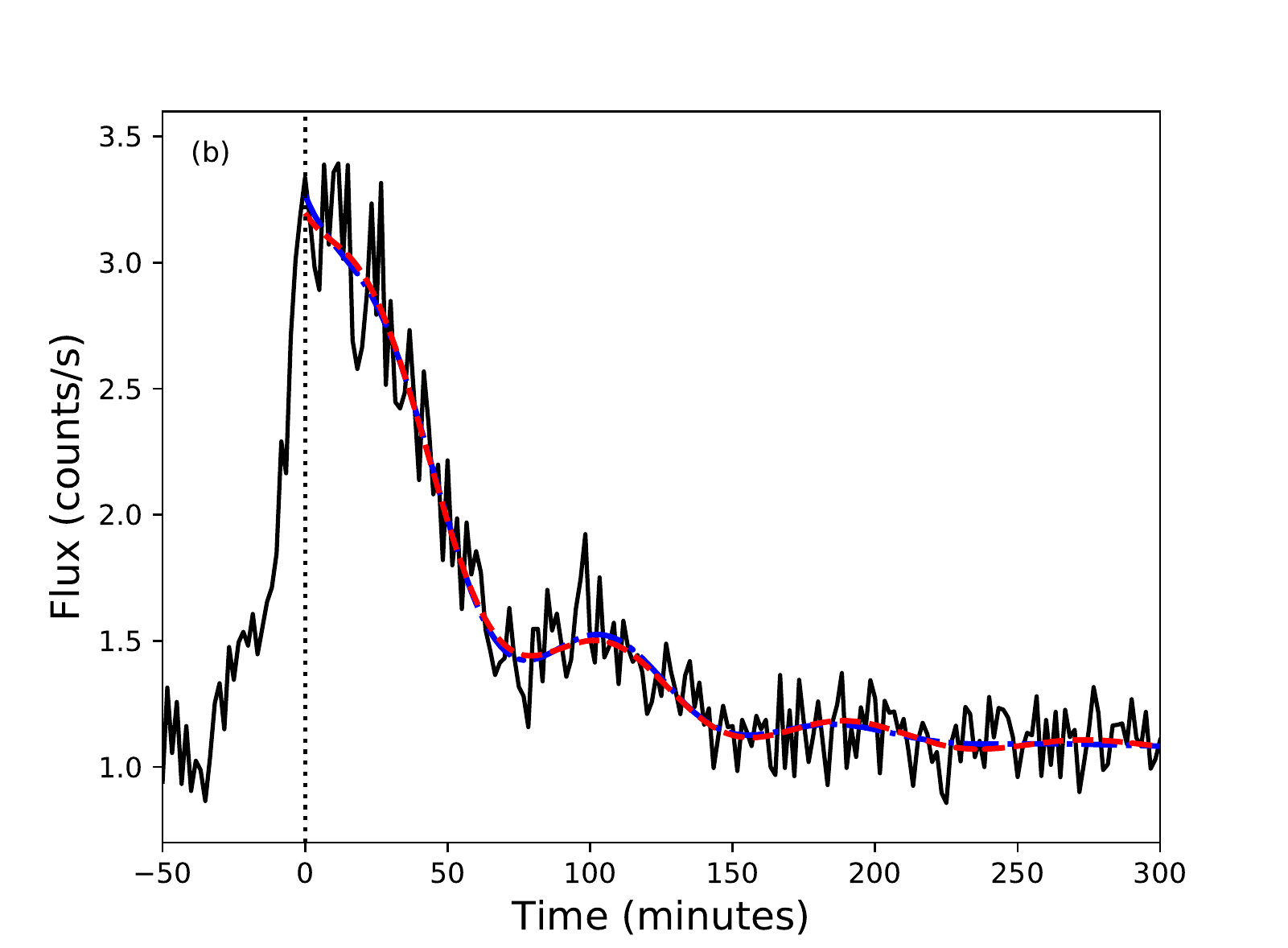}\\
   \caption{Panel (a): Flare lightcurve (black, solid) for low-energy-band data (0.2-1.0\,keV). Panel (b): Flare lightcurve (black, solid) for high-energy-band data (1.0-12.0\,keV). Also plotted in both panels are fits to the lightcurves consisting of the sum of an exponential decay term and a decaying sinusoid. The red, dashed line shows the fit when the decay of the sinusoid was described by an exponential, while the blue, dot-dashed curved shows the fit when the decay of the sinusoid was described by a Gaussian. We note that the blue and red curves are almost identical and, therefore, difficult to distinguish. The vertical dotted line highlights $t=0$\,min, which was determined from the total energy band.}
              \label{figure[flare_split]}%
    \end{figure*}

\begin{figure*}
   \centering
    \includegraphics[width=0.45\textwidth]{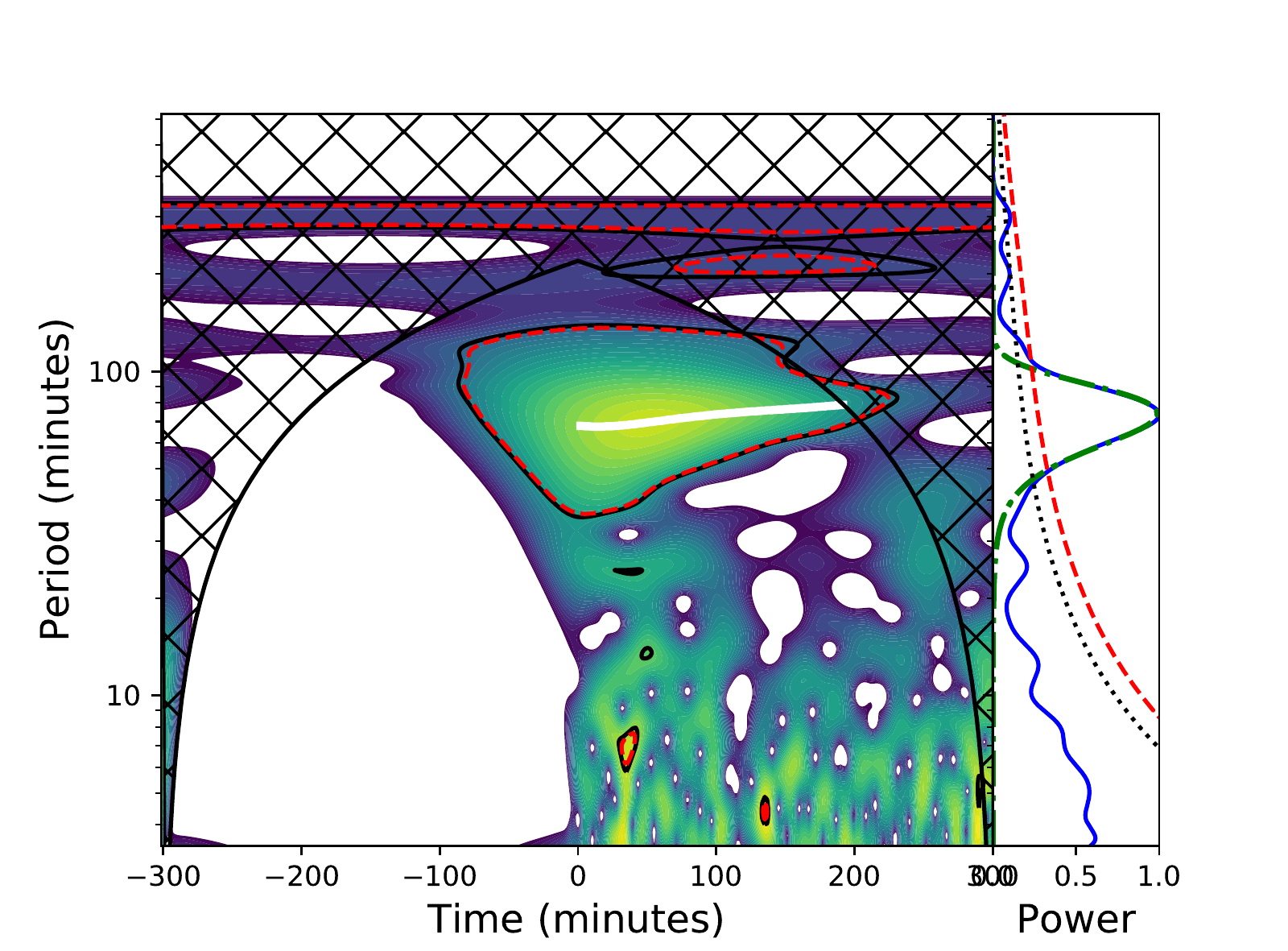}
    \includegraphics[width=0.45\textwidth]{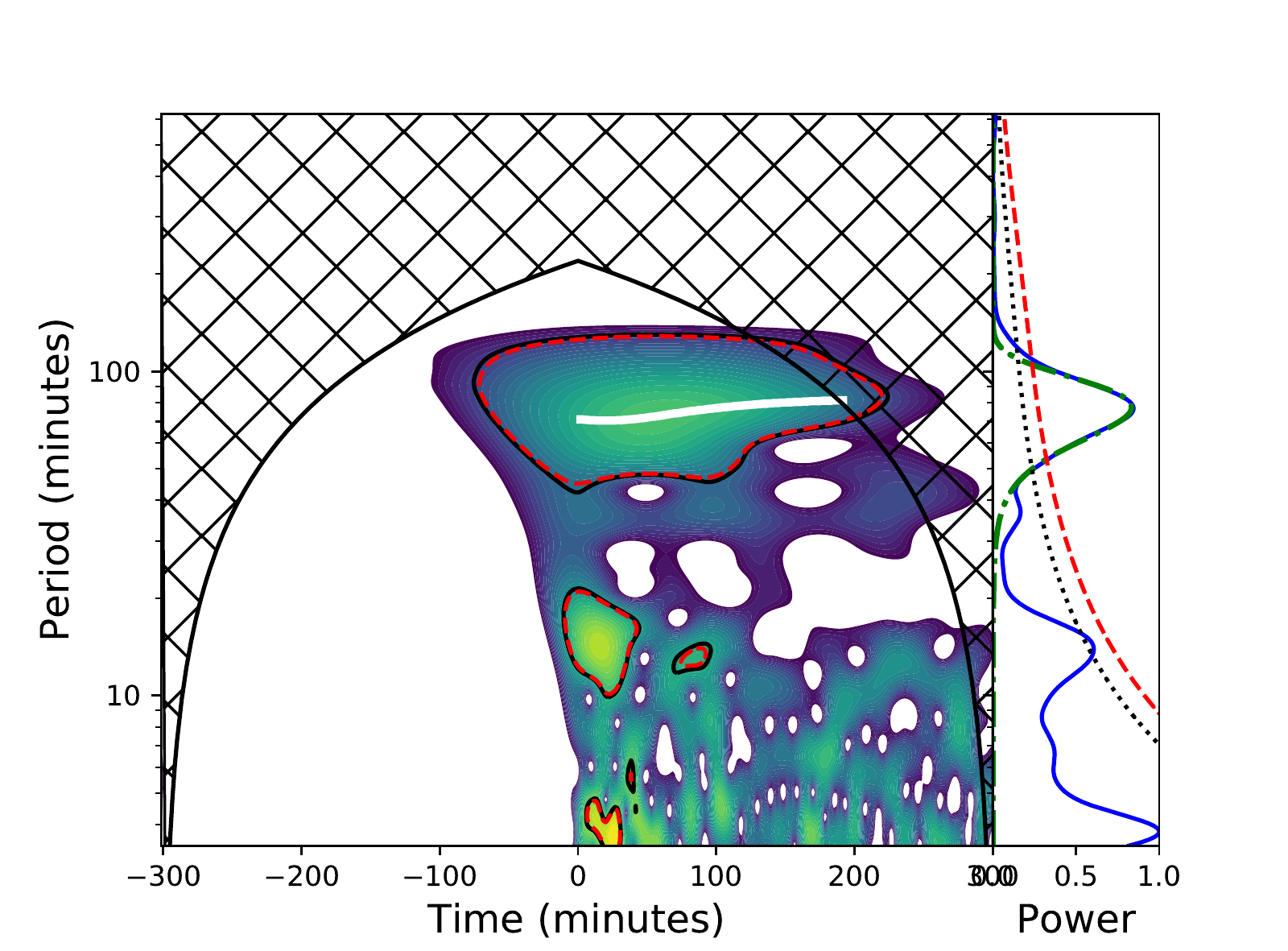}\\
   \caption{Left: Wavelet and global wavelet spectrum of detrended lightcuve for low-energy-band data (0.2-1.0\,keV). Right: Wavelet and global wavelet spectrum of detrended lightcurve for high-energy-band data (1.0-12.0\,keV). Contours are as described in Fig. \ref{figure[wavelet]}. The peak of the global spectrum has been fitted with a Gaussian (green, dot-dashed line).}
              \label{figure[wavelet_split]}%
    \end{figure*}

Figure \ref{figure[flare_split]} shows the flux observed when the data are split into two congruent energy bands, with the left-hand panel showing results for the low-energy band, which we initially focus on. Overplotted are the full-flare exponential and Gaussian fits. We note that $t=0\,\rm min$ corresponds to the peak flare intensity observed in the total ($0.2-12.0\,\rm keV$) energy band. The fitted curves are in good agreement with each other and the median and quartile uncertainties for the fitted parameters can be found in Table \ref{table[total_fits_bands]}. Once again fits using the exponentially decaying sinusoid were more stable, with over 3,000 of the 5,000 Monte Carlo fits failing for the Gaussian-decaying sinusoid. For the low-energy band both the Gaussian and exponential fits produced a periodicity of $73\pm2\,\rm min$. Although not shown here, fits to the low-energy band residuals obtained periods of $79\pm2$\,min and $77\pm2$\,min for the exponentially- and Gaussian-decaying sinusoids respectively, while the autocorrelation was also found to have a periodicity of $79\pm2\,\rm min$. The left-hand panel of Fig. \ref{figure[wavelet_split]} shows the wavelet spectrum for the low-energy band. Once again a significant feature is observed with a periodicity of approximately 70\,min and the ridge of maximum power evolves from $68\,\rm min$ at $t=0\,\rm min$ to $79\,\rm min$ at approximately $t=190\,\rm min$. A significant peak is also observed in the global wavelet spectrum and fitting a Gaussian curve to this peak reveals a periodicity of $74\pm16$\,min.

\begin{table*}\caption{Comparison of parameters obtained when fitting full flare for low- (0.2-1.0\,keV) and high-energy bands (1.0-12\,keV). The parameters in the top half were found when an exponentially-decaying sinusoid was included in full-flare fit, and the parameters in the bottom half were obtained when a Gaussian-decaying sinusoid was included in full-flare fit.}\label{table[total_fits_bands]}
\centering
\begin{tabular}{c|cc|cc}
  \hline
 & \multicolumn{2}{c}{Low energy band} & \multicolumn{2}{c}{High energy band} \\
Parameter & Histogram fit  & Median & Histogram fit  & Median \\
  \hline
$A_0$ (counts/s) & $2.81\pm0.15$ & $2.81\pm0.10$ & $2.40\pm0.10$ & $2.40\pm0.07$ \\
$t_0$ (min) & $56\pm4$ & $56\pm3$ & $51\pm3$ & $51\pm2$\\
$C$ (counts/s) & $3.30\pm0.03$ & $3.30\pm0.02$ & $1.08\pm0.01$ & $1.08\pm0.01$ \\
$A_e$ (counts/s) &  $0.627\pm0.014$ & $0.627^{+0.009}_{-0.010}$ & $0.623\pm0.0172$ & $0.622_{-0.13}^{+0.12}$\\
$B_e$ (counts/s) &  $23\pm7$ & $22^{+4}_{5}$ & $-33\pm20$ & $-45^{-14}_{+13}$\\
$\tau_e$ (min) &  $71\pm15$ & $71^{+10}_{-9}$ & $92\pm16$ & $93^{+12}_{-10}$ \\
$P_e$ (min) & $73\pm3$ & $73\pm2$ & $82\pm3$ & $82\pm2$ \\
$\phi_e$ (radians) & $2.8\pm0.2$ & $2.8\pm0.1$ & $4.0\pm0.2$& $4.0\pm0.2$ \\
\hline
$A_0$ (counts/s) & $2.69\pm0.12$ & $2.69\pm0.08$ & $2.39\pm0.08$& $2.39\pm0.05$\\
$t_0$ (min) & $60\pm4$ & $60\pm3$ & $52\pm3$& $52\pm2$\\
$C$ (counts/s) & $3.28\pm0.03$ & $3.28\pm0.02$ & $1.08\pm0.02$& $1.08\pm0.01$\\
$A_g$ (counts/s) &  $0.7\pm0.2$ & $0.8^{+0.3}_{-0.2}$ & $0.32\pm0.06$& $0.33^{+0.09}_{-0.04}$\\
$B_g$ (counts/s) &  $-39\pm80$ & $-54^{+44}_{-59}$ & $28\pm37$& $11\pm62$ \\
$\tau_g$ (min) &  $95.8\pm38$ & $96^{+28.0}_{-23}$ & $83\pm39$ & $83^{+31}_{-22}$\\
$P_g$ (min) & $73\pm3$ & $73\pm2$ & $82\pm4$& $82\pm2$\\
$\phi_g$ (radians) & $2.9\pm0.2$ & $2.9\pm0.1$ & $4.0\pm0.2$& $4.0\pm0.2$\\
  \hline
\end{tabular}
\end{table*}

Figure \ref{figure[flare_split]} also shows the full flare fits to the flux observed in the high-energy band and the fitted values are given in Table \ref{table[total_fits_bands]}. As with the other fits the exponentially-decaying sinusoid produces more stable fits than the Gaussian-decaying sinusoid. However, the fits for the Gaussian-decaying sinusoids for the high-energy band were more stable than for the low-energy band (with less than 1,500 fits failing). The period obtained from fitting the exponentially-decaying sinusoid is $82\pm2$\,min. Once again the fitted periodicities are in good agreement with those found in the detrended residuals and in the autocorrelation, which are not shown here. It is also in good agreement with the results of the wavelet transform, shown in Fig. \ref{figure[wavelet_split]}: a feature above the 99\% significance is observed with a periodicity of approximately $80\,\rm min$. The ridge of maximum power evolves from $70\,\rm min$ at $t=0\,\rm min$ to $82\,\rm min$ at approximately $190\,\rm min$ and fitting a Gaussian to the global spectrum indicates the significant peak has a periodicity of $77\pm17\,\rm min$.

Since the fits using the exponentially decaying sinusoid are more stable we now concentrate on these results. The period of the QPP found in the high-energy band is significantly longer than the QPP period detected in the low-energy band (with a difference of more than $3\sigma$). The phase is also significantly different (more than $5\sigma$).  However, we note that the difference in QPP period obtained from the wavelet spectra are less substantial and, because the peaks are broad, not significant. There is a possibility that the observed discrepancies in phase and period may be artefacts of the analysis procedure, and, more specifically, caused by fact that the parameters used to describe the underlying flare ($A_0$, $t_0$, and $C$) were different in the two energy bands. In particular, 2D histograms of the Monte Carlo parameters obtained for the full flare fits (see Appendix \ref{section[2d_hist]}) imply this is potentially related to the different values of $A_0$ obtained since $A_0$ is anti-correlated with both $P_e$ and $\phi_e$. In other words, in the Monte Carlo simulations a decrease in $A_0$, as is observed when we move from the low- to high-energy bands, leads to an increase in both period and phase, as required to remove the discrepancy. However, the data space of values of $A_0$, $P_e$, and $\phi_e$ sampled in the two sets of Monte Carlo simulations in the two bands are well separated, meaning that values of $A_0$ in the high-energy band that are sufficiently high to remove the discrepancy in period and phase are very unlikely to be the best fitting parameters. Furthermore, we note that the values of $A_0$ observed in each energy band are related to the underlying photon count rate of that band and so there is no physical reason to expect that these should be the same in both energy bands. This adds confidence to the fact that the discrepancies in the observed periods and phases are real. In addition, we note that although periods and phases are also correlated with $t_0$ the correlation is perpendicular to that required to explain the discrepancy. No correlation is observed with $C$.
    
\begin{figure}
   \centering
    \includegraphics[width=0.4\textwidth]{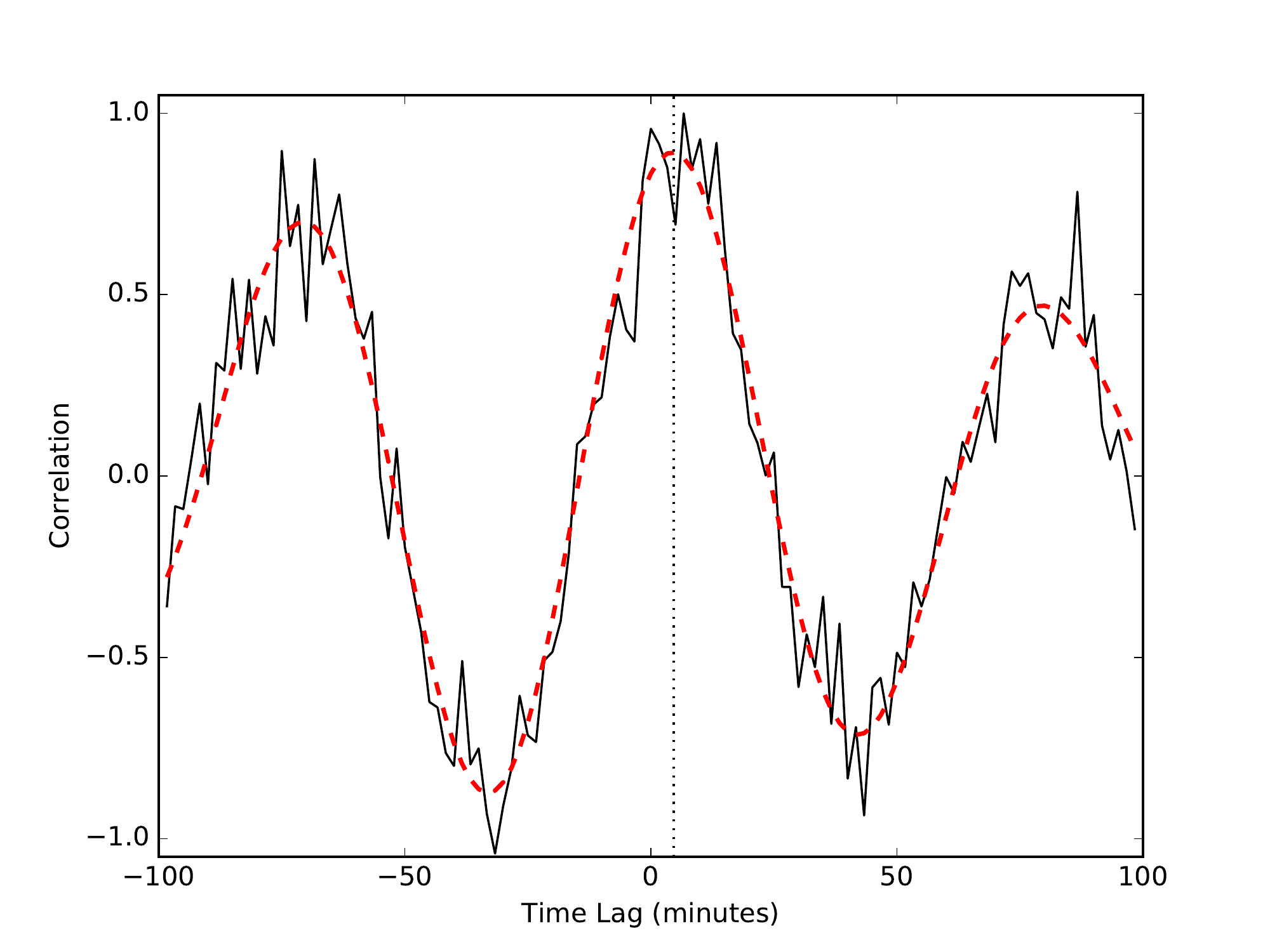}\\
   \caption{Cross correlation between residuals observed in low- and high-energy bands (0.2-1.0\,kev and 1.0-12.0\,kev). The red-dashed line is a sinusoidal fit to the data with a Gaussian decay. The black dotted line indicates the lag of the peak of this fit i.e. $4.7\pm1.3\,\rm min$. }
              \label{figure[cross_correlation]}%
    \end{figure}

Figure \ref{figure[cross_correlation]} shows the cross-correlation of the low- and high-energy-band residuals. The peak of the cross correlation is offset from zero. To determine the significance of this offset 5,000 Monte Carlo simulations were used, similar to those described in Sect. \ref{section[analysis]}, where a Gaussian-decaying sinusoid was fitted to each simulated cross-correlation. This enabled a histogram of the determined offset to be produced, which was well described by a Gaussian. The median value of the offset was found to be $4.7\pm1.3$\,min, where the uncertainties are given by the quartile range. The offset is, therefore, significant at more than a $3\sigma$ level.

\subsection{Comparison of results when splitting the data into separated energy bands} \label{section[separate]}
\begin{figure*}
   \centering
    \includegraphics[width=0.45\textwidth]{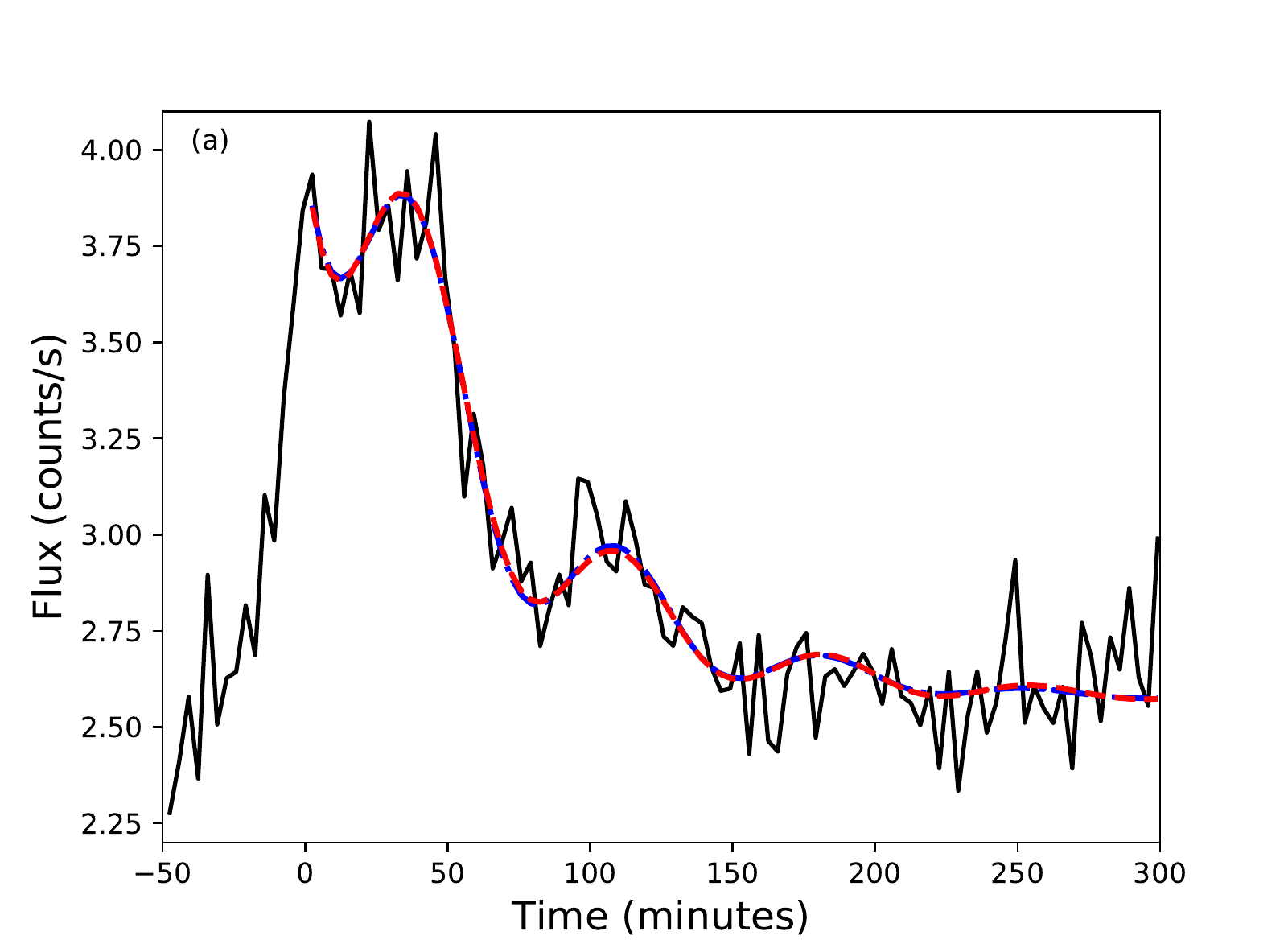}
    \includegraphics[width=0.45\textwidth]{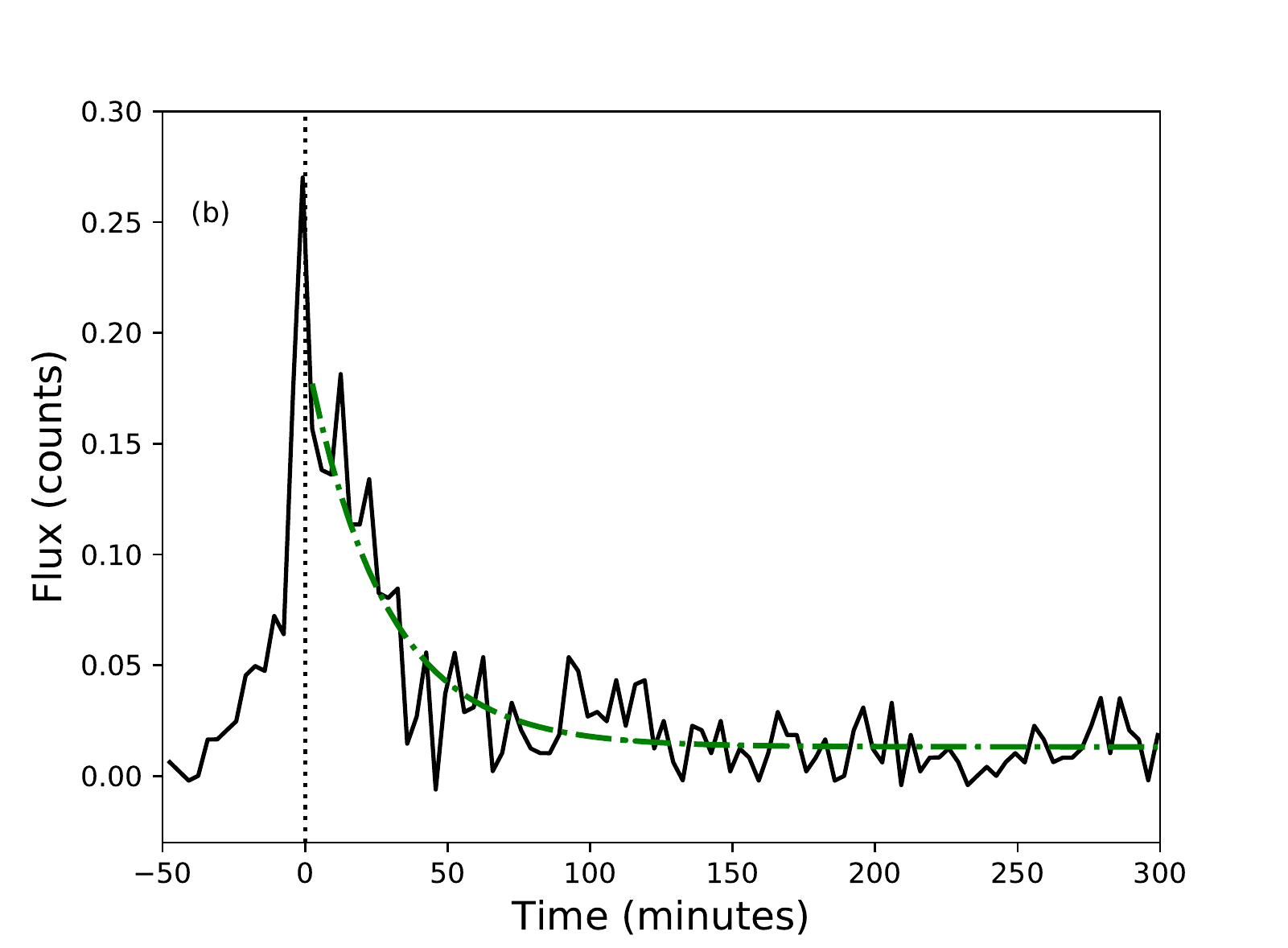}\\
   \caption{Panel (a): Flare lightcurve (black, solid) for data observed between 0.5 and 1.0\,keV. The red, dashed line shows the fit when the decay of the sinusoid was described by an exponential, while the blue, dot-dashed curved shows the fit when the decay of the sinusoid was described by a Gaussian. We note that the blue and red curves are almost identical and, therefore, difficult to distinguish. Panel (b): Flare lightcurve (black, solid) for data observed between 4.5 and 12.0\,keV. No good fit containing QPPs was obtained for this data set and so the green dot-dashed line shows the best fitting exponential decay as described by Eq. \ref{eqn: decay}. }
              \label{figure[flare_split_gap]}%
    \end{figure*}

Figure \ref{figure[flare_split_gap]} shows the flux observed in the $0.5-1.0$\,keV energy band, with the full-flare fits. The Gaussian and exponential fits are in good agreement with each other and the values of the full-flare exponential fitted parameters can be found in Table \ref{table[total_fits_split_bands]}. The period obtained was $71\pm2$\,min. Also plotted in Fig. \ref{figure[flare_split_gap]} is the flux observed in the higher-energy band, covering the range $4.5-12.0\,\rm keV$. The flux values here are very low and it was not possible to obtain a robust full-flare fit and so Fig. \ref{figure[flare_split_gap]} shows only the exponential decay fit to the flare, given by Eq. \ref{eqn: decay}. The fitted parameters for this fit can be found in Table \ref{table[total_fits_split_bands]}. 

\begin{table*}\caption{Comparison of parameters obtained when fitting summations of Eq. \ref{eqn: decay} and either Eq. \ref{eqn: oscillations exp} or Eq. \ref{eqn: oscillations gauss} to flux for low-energy band and just Eq. \ref{eqn: decay} to the high-energy band.}\label{table[total_fits_split_bands]}
\centering
\begin{tabular}{c|cc|cc}
  \hline
 & \multicolumn{2}{c}{Low energy band} & \multicolumn{2}{c}{High energy band} \\
Parameter & Histogram fit  & Median & Histogram fit  & Median \\
  \hline
$A_0$ (counts/s) & $1.92\pm0.13$ & $1.93_{-0.08}^{+0.09}$ & $0.18\pm0.02$ & $0.18\pm0.01$ \\
$t_0$ (min) & $55\pm5$ & $55_{-3}^{+4}$ & $27\pm4$ & $27\pm2$\\
$C$ (counts/s) & $2.58\pm0.03$ & $2.58\pm0.02$ & $0.013\pm0.001$ & $0.013\pm0.001$ \\
$A_e$ (counts/s) &  $0.61\pm0.02$ & $0.61\pm0.01$ & n/a & n/a\\
$B_e$ (counts/s) &  $3\pm13$ & $2^{+8}_{11}$ & n/a & n/a\\
$\tau_e$ (min) &  $71\pm17$ & $71^{+11}_{-13}$ & n/a & n/a\\
$P_e$ (min) & $71\pm3$ & $71\pm2$ & n/a & n/a \\
$\phi_e$ (radians) & $2.5\pm0.3$ & $2.5\pm0.2$ & n/a & n/a \\
  \hline
\end{tabular}
\end{table*}
    
\begin{figure*}
   \centering
    \includegraphics[width=0.45\textwidth]{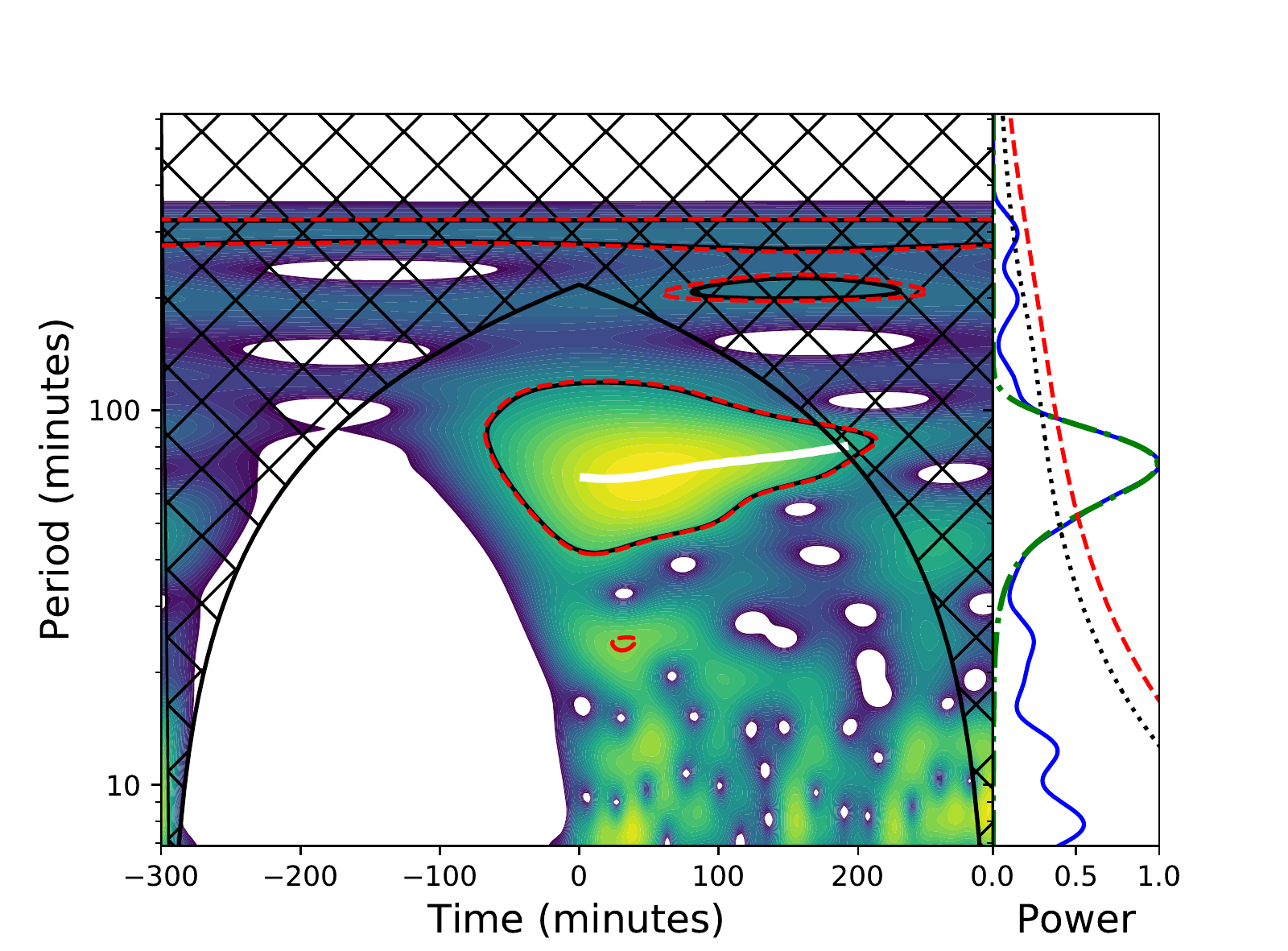}
    \includegraphics[width=0.45\textwidth]{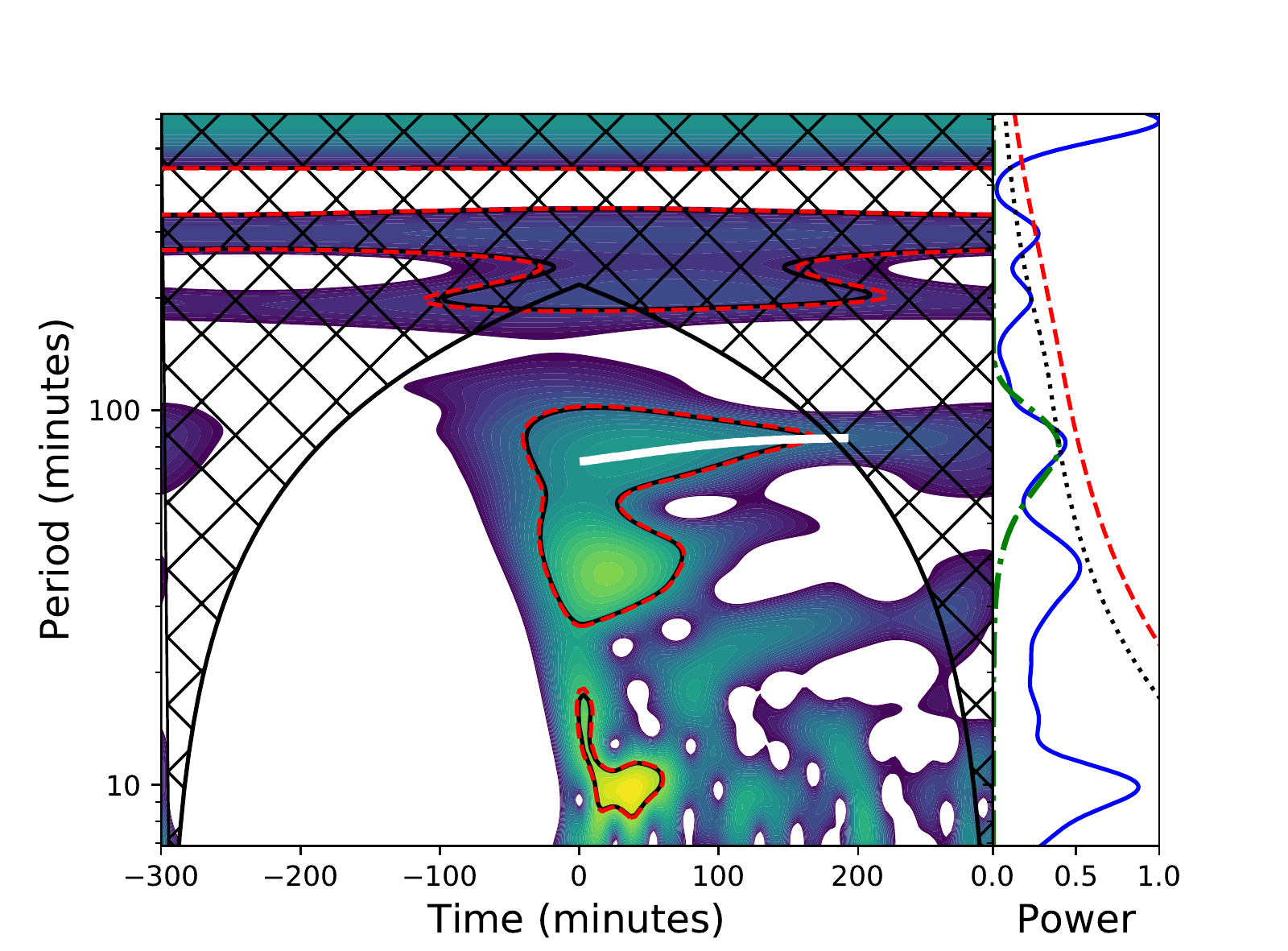}\\
   \caption{Left: Wavelet and global wavelet spectrum of detrended lightcurve for data in 0.5-1.0\,keV energy range. Right: Wavelet and global wavelet spectrum of detrended lightcurve for data in 4.5-12.0\,keV energy range. In both panels, contours and lines are as described in Fig. \ref{figure[wavelet]}. }
              \label{figure[wavelet_split_gap]}%
    \end{figure*}

Figure \ref{figure[wavelet_split_gap]} shows the wavelet transforms of the residuals for both the $0.5-1.0$\,keV range and the $4.5-12.0$\,keV range. The $0.5-1.0$\,keV range shows a significant periodicity of around 70\,min. The ridge of maximum power initially decreases in period before reaching a minimum of $66\,\rm min$ at $t=15\,\rm min$. The ridge then evolves to higher periods, reaching $80\,\rm min$ at $t=190\,\rm min$. The global wavelet spectrum also shows a significant peak and fitting a Gaussian curve to the peak indicates a period of $72\pm17$\,min, where the uncertainties are given by the width of the peak in the global power spectrum, which is broad. The wavelet of the residuals of the flux observed in the  $4.5-12.0$\,keV energy band also shows a significant peak, which appears to be split into two bands, one with a periodicity of around 80\,min and a second, short-lived periodicity at around 40\,min. The ridge of maximum power for the period around $80\,\rm min$ evolves from $74\,\rm min$ at $t=0\,\rm min$ to $84\,\rm min$ at $t=190\,\rm min$. We note that the 40\,min periodicity has a higher power but the peak is short lived. In the global spectrum the peak at around 80\,min is significant above the 99\% level only if the traditional significance levels are used. When the modifications recommended by \citet{2016ApJ...825..110A} are incorporated this peak is not significant at the 99\% level. Nevertheless a Gaussian curve was fitted to this which was found to have a maximum at $80\pm20$\,min, where the uncertainties are given by the width of the peak. The secondary peak is above neither set of significance levels. 

\begin{figure}
   \centering
    \includegraphics[width=0.4\textwidth]{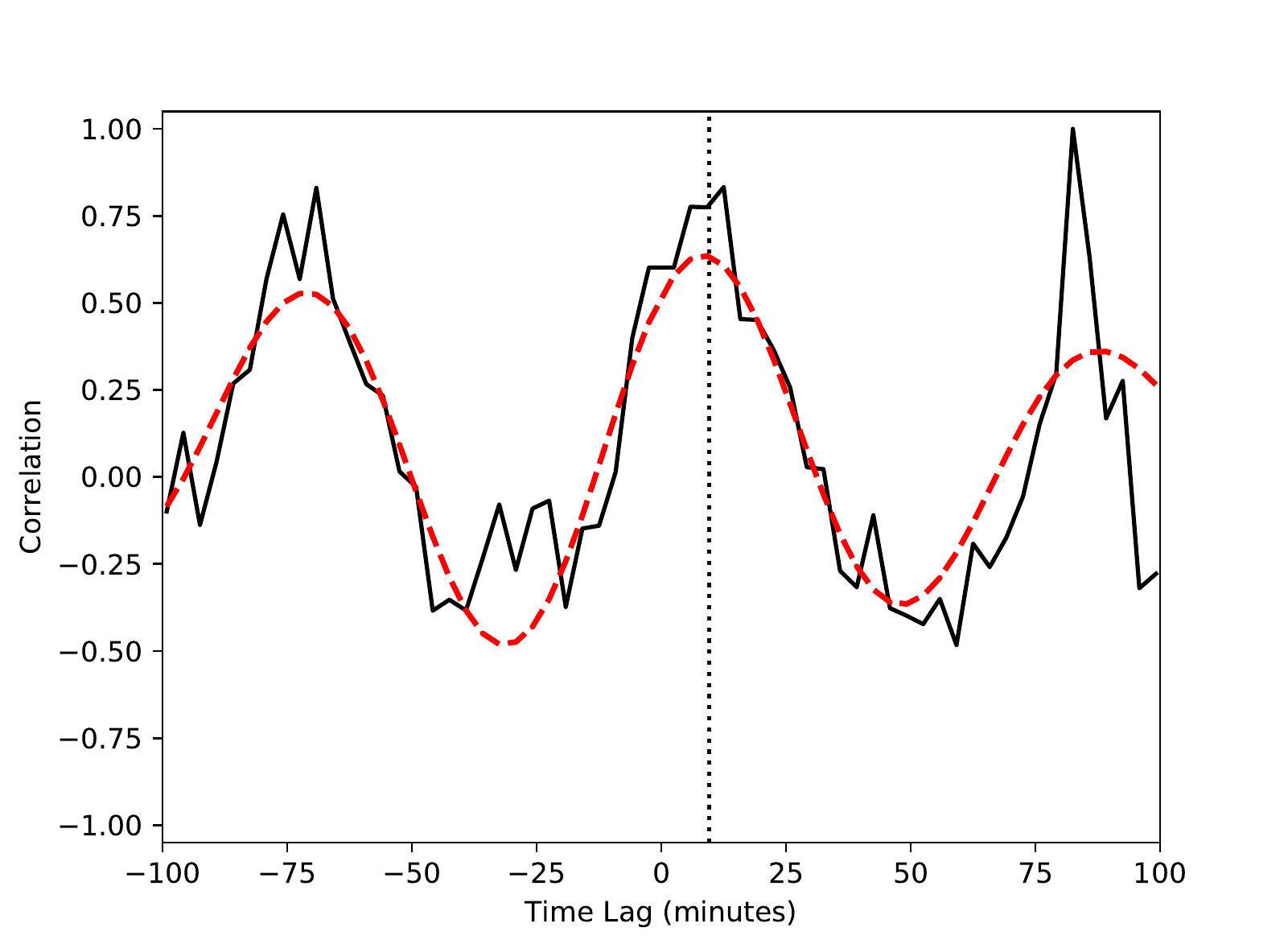}\\
   \caption{Cross-correlation between detrended flares observed in two different energy bands (0.5-1.0\,kev and 4.5-12.0\,kev). The red-dashed line is a sinusoidal fit to data with a Gaussian decay. The black dotted line indicates the lag of the peak of this fit i.e. $10\pm3\,\rm min$. }
              \label{figure[cross_correlation_split_gap]}%
    \end{figure}

Figure \ref{figure[cross_correlation_split_gap]} shows the cross-correlation of the $0.5-1.0$\,keV and $4.5-12.0$\,keV residuals. Once again there is an offset in the peak from zero. Using Monte Carlo simulations this offset is found to be $10\pm3\,\rm min$. This is a larger offset than seen in the congruent energy bands, however, we note that the uncertainties are far larger because of the low flux in the $4.5-12.0$\,keV energy band. The period of the cross correlation is $81\pm3$\,min. 

\section{Discussion}\label{section[discussion]}


The similarity in appearance between the QPPs observed in this flare and those QPPs in \textit{Kepler} white light flares studied in \citet{2016MNRAS.459.3659P} is interesting.  Indeed, we are able to successfully fit the same combined QPP and flare model described by \citet{2015ApJ...813L...5P, 2016MNRAS.459.3659P}. The periodicity observed here, of $76\pm2$\,min, is within the range observed by \citet{2016MNRAS.459.3659P} of $9-90$\,min. Although solar QPPs with periods of the order of tens of minutes have been observed \citep{2013A&A...555A..55Z}, periods less than 10\,min are far more common \citep[see e.g.][for recent surveys]{2016ApJ...833..284I, 2017A&A...608A.101P}. This is likely to be a selection effect as solar flares tend to be far less energetic, and therefore shorter lived, than stellar flares. 

Using a subset of the \textit{Kepler} QPP flares with stable Gaussian-decaying oscillations \citet{2016MNRAS.459.3659P} found an empirical relationship between period, P, and damping time, $\tau$:
\begin{equation}\label{eqn[pugh2016]}
\ln\tau=(1.1\pm0.01)\ln P-{0.38\pm0.01}.
\end{equation}
Using Eq. \ref{eqn[pugh2016]} and the period found here when fitting a Gaussian-decaying sinusoid to the residuals ($P_g=78.1^{+1.1}_{-1.2}$\,min) we would expect to observe a damping time of $87\pm4$\,min, which is in reasonable agreement with the value of $\tau_g=79^{+20}_{-14}$\,min found here, although we recall that an exponentially decaying sinusoid produces a far more stable fit to the QPPs. We note here that \citet{2016MNRAS.459.3659P} found no significant correlation between periods and damping times for those QPPs with stable exponentially-decaying oscillaions, of which there were only five in the sample. 

By fitting a exponentially-decaying sinusoid to residuals once a flare, observed by XMM-Newton, had been detrended using Empirical Mode Decomposition, \citet{2016ApJ...830..110C} also found an empirical relationship between period and damping time of
\begin{equation}\label{eqn[cho2016]}
\tau=(1.70\pm1.13)P^{0.98\pm0.05}.
\end{equation}
Furthermore, \citet{2016ApJ...830..110C} found that this relation was consistent with the relationship, obtained using the same techniques, between periods and damping rates of QPPs observed in solar flares, suggesting a common physical origin. Using Eq. \ref{eqn[cho2016]} and the period obtained by fitting an exponentially-decaying sinusoid to the residuals ($P_e=78.8_{-0.8}^{+0.9}$\,min) we would expect a damping time of $122\pm86$\,min, which is poorly constrained because of the uncertainty in the parameters fitted by \citet{2016ApJ...830..110C}. This can nevertheless be compared to the value of $95\pm7$\,min obtained here.

A similar relationship is found for transverse waves in the solar corona by \citet{2013A&A...552A.138V}:  
\begin{equation}\label{eqn[verwichte2013]}
\log_{10}\tau=(0.44\pm0.31)+(0.94\pm0.12)\log_{10}P.
\end{equation}
Their relationship predicts an exponential damping time of $167\pm61\,\rm min$, which again is longer than the damping time observed here but is too poorly constrained to make any definitive conclusions. We also note that the majority of the oscillations included in \cite{2013A&A...552A.138V} have periods less than $17\,\rm min$. Nevertheless, it appears that the values of period and damping rates obtained in this work are consistent, at least, with previous solar and stellar studies.

By splitting the data into different energy bands we have observed that a statistically significant QPP signal is present at both low and high energies. However, when the data are fitted, the period in the high-energy band was found to be significantly longer than the period in the low-energy band and the phase of the signals in the two energy bands was significantly different, with the high-energy band leading the low-energy band. This could indicate that processes are occurring in more than one plasma structure, such as different loops with different resonant periods. 
 
Alternatively, the phase shift between QPP signals in the two congruent energy bands could suggest the flares, or even the QPPs themselves, are also subject to the Neupert effect \citep{1968ApJ...153L..59N}, which is the empirical tendancy for high-energy X-ray emission observed during a flare to coincide with the temporal derivative of the lower-energy X-ray emission. Since we have taken the same time for $t=0\,\rm min$ for each energy band, the presence of the Neupert effect in the flares themselves could modify the phase of the observed QPPs in the two energy bands. However, it is also possible for the Neupert effect to materialise in the QPP signal itself:  the temperature dependence of cooling processes can introduce delays in the peak times of energy bands dominated by thermal effects compared to higher-energy bands \citep[e.g.][]{2011SSRv..159..107H}, resulting in, for example, a QPP signal in soft X-rays that lags a QPP signal observed in hard X-rays \citep{2012ApJ...749L..16D}. Although both channels investigated here (with an upper limit of $12\,\rm keV$) are likely to be dominated by thermal emission, we note that the low-energy band lags behind the high-energy band. The presence of the Neupert effect in the observed QPPs may imply that the QPPs are caused by the modulation of the propagation speeds or acceleration of charged particles, rather than the direct modulation of the X-ray intensity.  Such a modulation could occur, for example, because of leakage of MHD waves from neighbouring structures that trigger periodic reconnection \citep{2006A&A...452..343N}. It is not clear though that the presence of a Neupert effect can explain the different periodicities observed. However, we note that the observed difference in period is far less significant than the observed difference in phase and the difference in periods observed in the wavelet spectra is not significant. Since the ridge of maximum power in the wavelet appears to evolve, with period increasing as a function of time, for each energy band it is possible that the QPP has a non-stationary period and by representing it as a stationary QPP in the models (Eqs. \ref{eqn: oscillations exp} and \ref{eqn: oscillations gauss}) we are underestimating the uncertainties associated with the period.

\section{Summary}\label{sec:summary}
We find statistically significant QPPs in the soft X-ray emission of the solar analogue, EK Dra during a flare. The decay phase of the flare is well-described by an exponential decay and the QPPs are best described by an exponentially-decaying sinusoid. This signal is initially found in the total-energy band, which covers the range $0.2-12.0\,\rm keV$, and is found to have a periodicity of $76\pm2\,\rm min$. The QPPs are also detected in two smaller congruent energy bands, namely $0.2-1.0\,\rm keV$ and $1.0-12.0\,\rm keV$. When models are fit to the data, the signals in these two bands were found to differ in period, by more than $3\sigma$, and in phase, by $9\sigma$. However, we note that, because the peaks in the wavelet spectra are broad, the  periods obtained from the wavelet spectra are not statistically significantly different. Nevertheless, the cross-correlation reveals a significant offset from zero and the high-energy band is found to lead the low-energy band, which is consistent with the Neupert effect, suggesting the QPPs are caused by the modulation of the propagation speeds or acceleration of charged particles. However, an alternative explanation is that we are observing processes in two different plasma structures. Finally, the appearance and properties of the QPPs studied here are consistent with those observed previously in both solar and stellar flares, hinting at that the same physics may link solar and stellar flares.

\begin{acknowledgements} 
AMB \& AT thank the Royal Astronomical Society for funding AT's summer project. AMB acknowledges
support from the Royal Society International Exchanges grant IEC\textbackslash R2\textbackslash 170056 and the International Space Science Institute for the team "Quasi-periodic Pulsations in Stellar Flares: a Tool for Studying the Solar-Stellar Connection." This work has made use of data from the European Space Agency (ESA) mission
{\it Gaia} (\url{https://www.cosmos.esa.int/gaia}), processed by the {\it Gaia}
Data Processing and Analysis Consortium (DPAC,
\url{https://www.cosmos.esa.int/web/gaia/dpac/consortium}). Funding for the DPAC
has been provided by national institutions, in particular the institutions
participating in the {\it Gaia} Multilateral Agreement.   
\end{acknowledgements}

\bibliographystyle{aa} 
\bibliography{ekdra}

\appendix
\section{Histograms for fits to total-energy band}\label{section[histograms]}

\begin{figure*}
   \centering
    \includegraphics[width=0.9\textwidth, trim=0cm 9cm 0cm 0cm, clip=true]{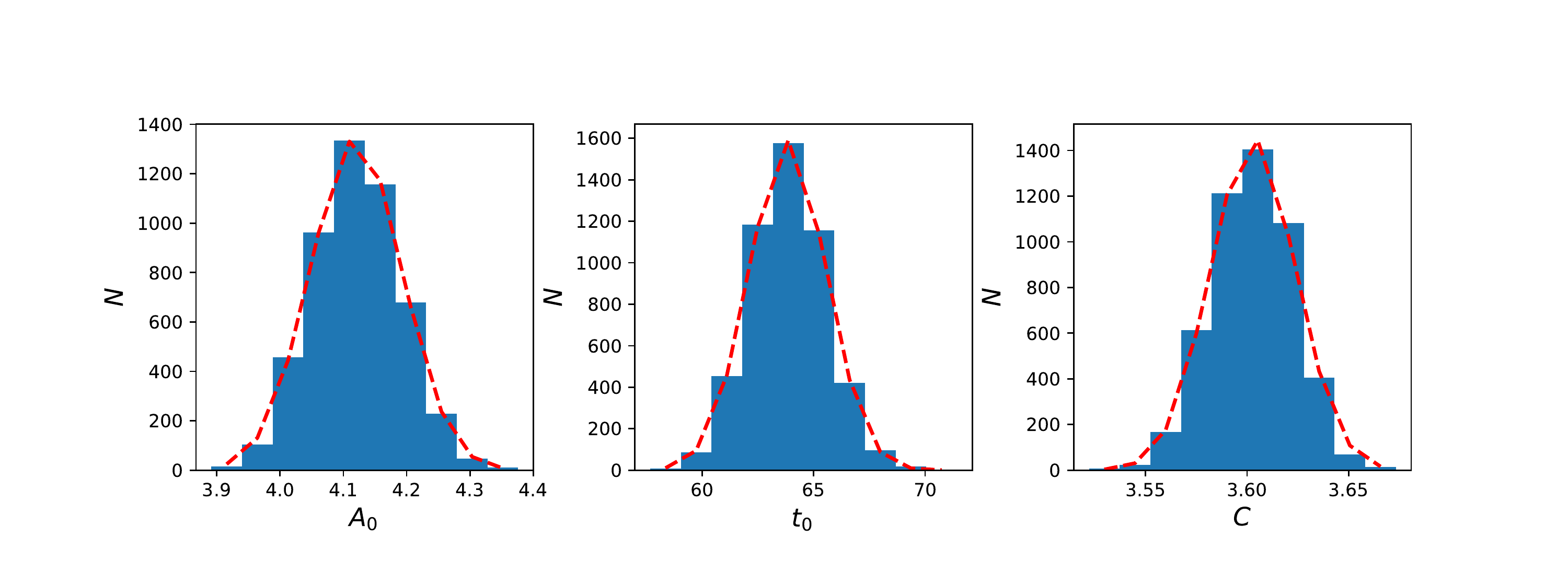}\\
   \caption{Histograms used to determine quality of fits and uncertainties on fitted parameters when fitting Eq. \ref{eqn: decay} to total-energy band count rates as shown in panel (a) of Fig. \ref{figure[flare]}. The red-dashed curve shows best fitting Gaussian.}
              \label{figure[hist_flare]}%
    \end{figure*}

\begin{figure*}
   \centering
    \includegraphics[width=0.9\textwidth]{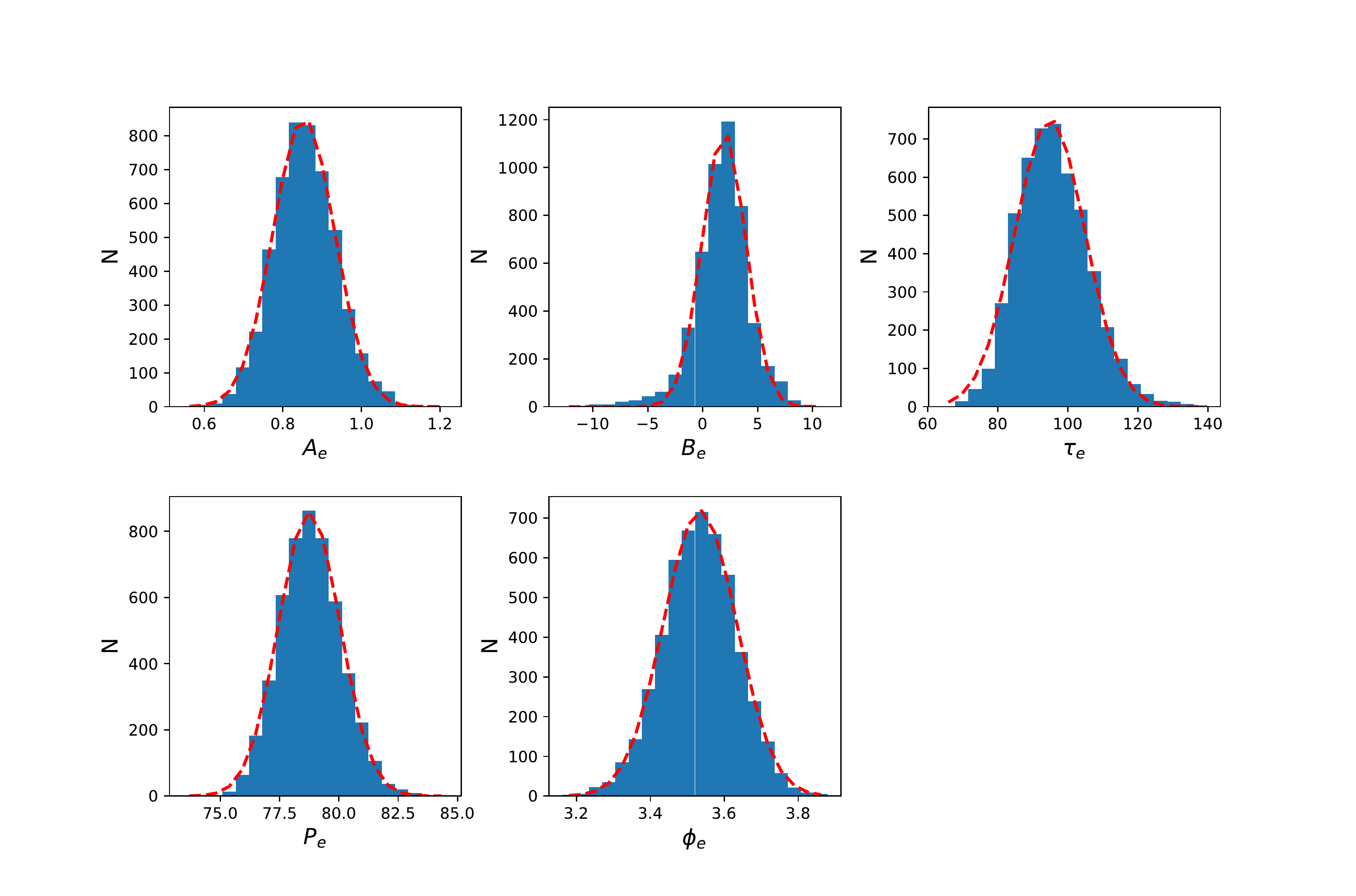}\\
   \caption{As in Fig. \ref{figure[hist_flare]} when fitting Eq. \ref{eqn: oscillations exp} to total-energy band residuals as shown in panel (b) of Fig. \ref{figure[flare]}.}
              \label{figure[hist_qpp_exp]}%
	\end{figure*}

\begin{figure*}
   \centering
    \includegraphics[width=0.9\textwidth]{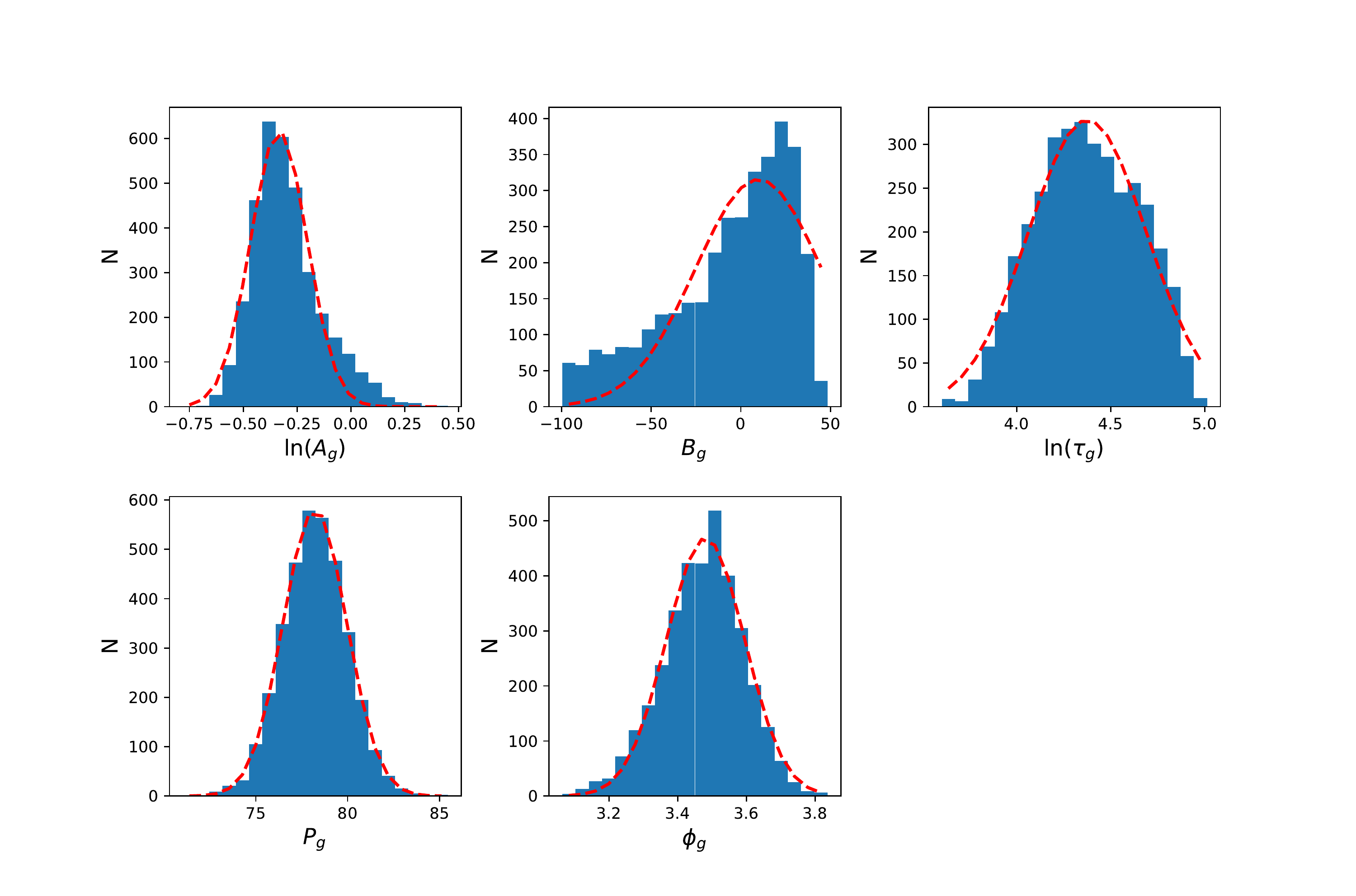}\\
   \caption{As in Fig. \ref{figure[hist_flare]} when fitting Eq. \ref{eqn: oscillations gauss} to total-energy band residuals as shown in panel (b) of Fig. \ref{figure[flare]}.}
              \label{figure[hist_qpp_gauss]}%
	\end{figure*}

\begin{figure*}
   \centering
    \includegraphics[width=0.9\textwidth]{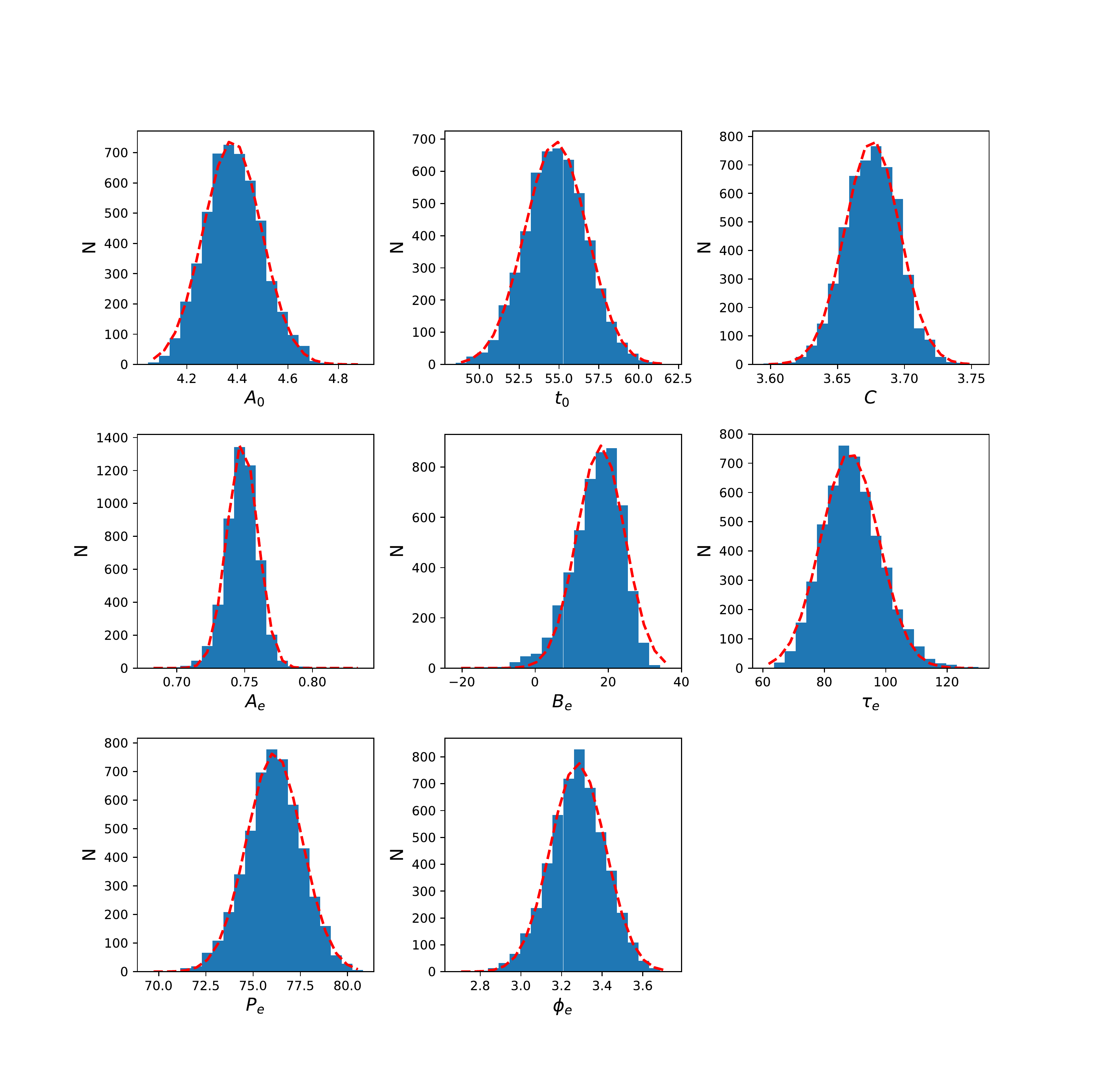}\\
   \caption{As in Fig. \ref{figure[hist_flare]} when fitting summation of Eq. \ref{eqn: decay} and \ref{eqn: oscillations exp} to total-energy band count rate as shown in panel (c) of Fig. \ref{figure[flare]}. }
              \label{figure[hist_full_exp]}%
	\end{figure*}

\begin{figure*}
   \centering
    \includegraphics[width=0.9\textwidth]{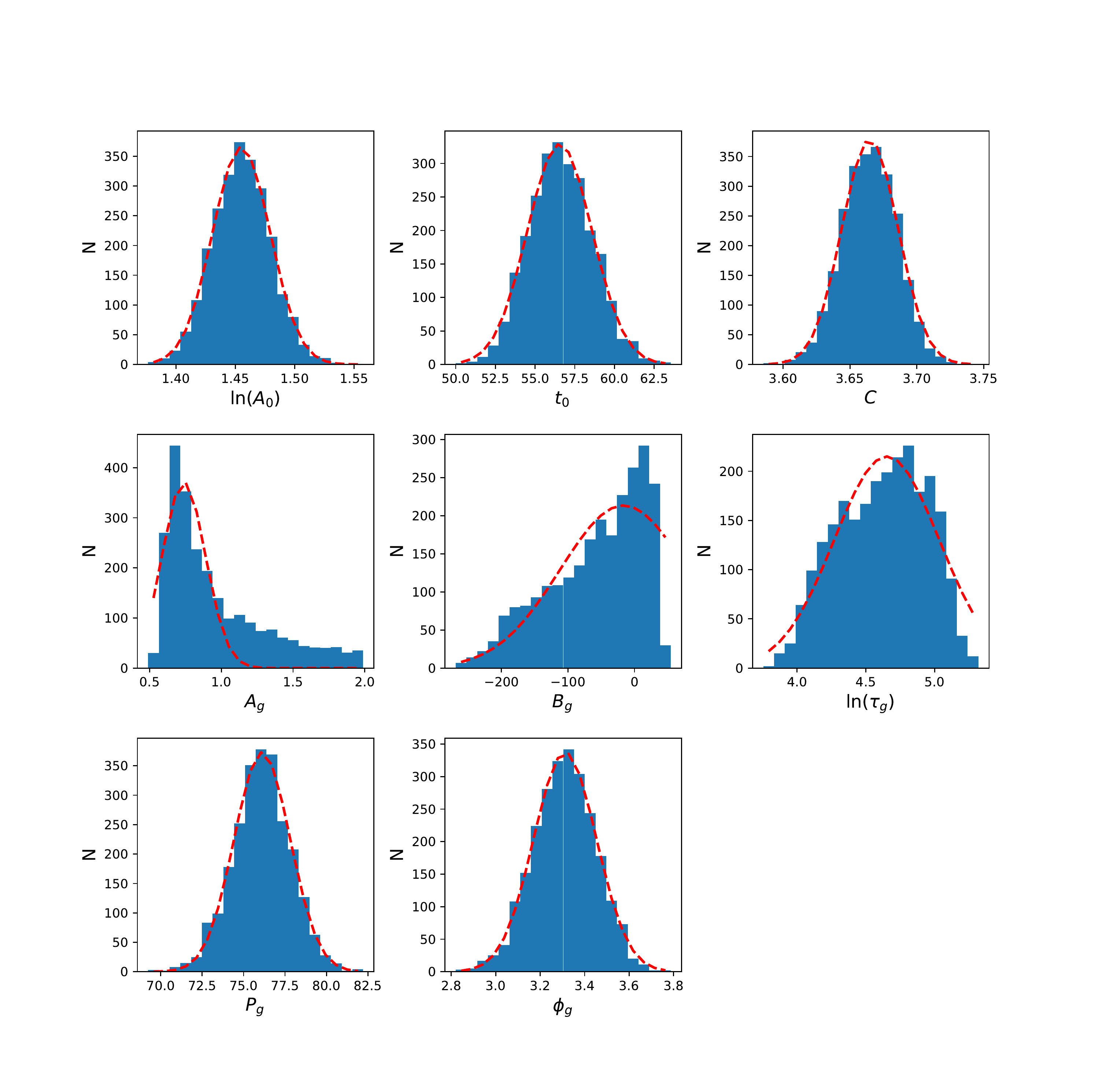}\\
   \caption{As in Fig. \ref{figure[hist_flare]} when fitting summation of Eq. \ref{eqn: decay} and \ref{eqn: oscillations gauss} to total-energy band count rate as shown in panel (c) of Fig. \ref{figure[flare]}.}
              \label{figure[hist_full_gauss]}%
	\end{figure*}

To determine both the quality of the fit to the data and to produce reliable uncertainty estimates we used Monte Carlo simulations where the fit was performed 5,000 times but each time a fit was performed the time series was modified by adding random numbers to each datum. The random numbers were taken from a normal distribution with a standard deviation given by the formal error on the data. Once the 5,000 fits had been performed we created histograms of the output parameters and these are shown in Figs. \ref{figure[hist_flare]} $-$ \ref{figure[hist_full_gauss]} for the various different fits performed in this article. As can be seen some of the parameters, such as those shown in Figs. \ref{figure[hist_flare]}, \ref{figure[hist_qpp_exp]}, and \ref{figure[hist_full_exp]}, are well constrained and the histograms produce symmetric normal distributions. However, when the Gaussian decaying sinusoid was fitted to the residuals and as part of the full flare fit some of the histograms are poorly represented by a Gaussian. We note that some of the parameters produce something closer to a normal distribution when the histograms are plotted in log space. This suggests that the fit is less stable and that an exponentially decaying sinusoid represents the data better than a Gaussian-decaying sinusoid. 

\section{Additional tests for significance in wavelet spectra} \label{sec: extra_wavelets}
\begin{figure}
   \centering
    \includegraphics[width=0.45\textwidth]{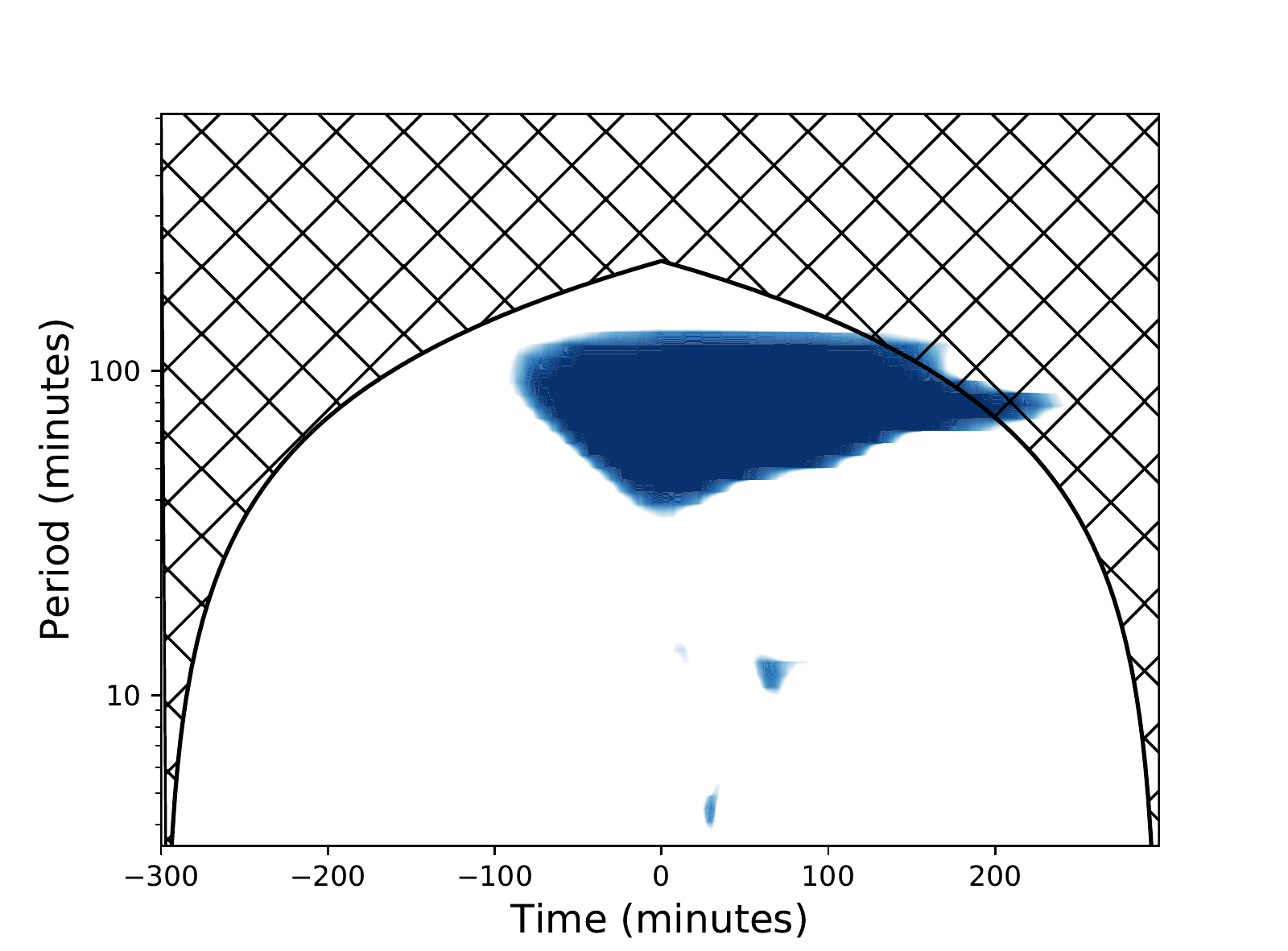}\\
   \caption{Wavelet spectrum of total-energy band, where significance levels were determined using Fisher Randomisation Test. Blue data points are all above the 99\% significance level as determined by this method.}
              \label{fig: FRT_wavelet}%
	\end{figure}

To test the reliability of the significance levels in the wavelet spectra we performed two additional tests. Firstly we performed a Fisher Randomisation Test \citep{fisher:1935} where the order of the data points in the residuals were randomly shuffled 5,000 times. Following each shuffle a wavelet spectrum was computed and stored. These stored wavelet spectra are used to create a cumulative histogram for each data point within the wavelet and, in turn, these histograms are used to determine the 99\% significance level, based on a H0 hypothesis that the data contain only noise. If a data point in the wavelet spectrum of the original residual time series is larger than 99\% of the wavelet spectra obtained from the shuffled residual time series that data point was highlighted as being significant. In the wavelet spectrum shown in Figure \ref{fig: FRT_wavelet} these significant data points are shown in blue. This spectrum was obtained using the total energy band and the similarity in the significance region with that found in Figure \ref{figure[wavelet]} is clear. Although only the total-energy band is shown here we note that the Fisher Randomisation Tests confirmed the significance of the signals observed in all frequency bands.

\begin{figure}
   \centering
    \includegraphics[width=0.45\textwidth]{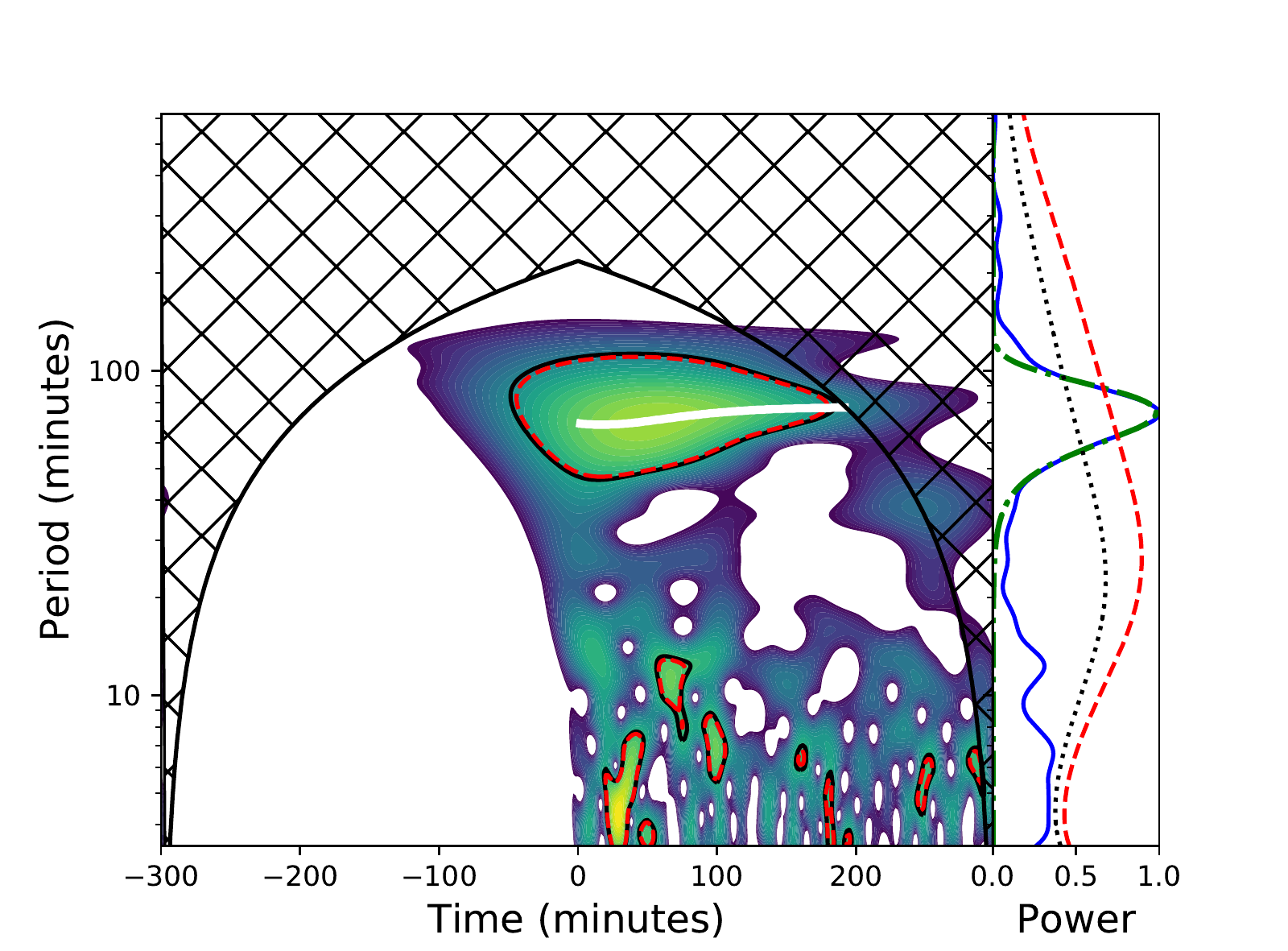}\\
   \caption{Wavelet spectrum of total-energy band, where significance levels were based on an assumption of red noise. Lines are as described in Figure \ref{figure[wavelet]}.}
              \label{fig: red_wavelet}%
	\end{figure}

Eliminating the underlying flare trend, here through the subtraction of Eq. \ref{eqn: decay}, removes the dominant background trend, which can also be thought of as a red noise signal in this instance. However, it is possible that some red noise remains in the residuals. We therefore also test the significance of the signal obtained in the wavelet spectrum based on an assumption of red noise. Figure \ref{fig: red_wavelet} shows the resulting wavelet spectrum obtained for the total-energy band. The main peak at approximately $80\,\rm min$ is still significant. Although only the total-energy band is shown here, the dominant QPP signal is found to be significant in all other energy bands based on this red noise assumption.

\section{2D histograms for full flare fit parameters obtained when examining the light curves from two congruent energy bands} \label{section[2d_hist]}
In Section \ref{section[congruent]} (specifically Figure \ref{figure[flare_split]} and Table \ref{table[total_fits_bands]}) it was demonstrated that the phase and period of the QPPs in two congruent energy bands differed substantially. However, there is a possibility that this is an artefact introduced by differences in the underlying flare shape. Here, we show 2D histograms of the QPP period ($P_e$) and phase ($\phi_e$) with the parameters that describe the underlying flare shape ($A_0$, $t_0$, and $C$) obtained in the Monte Carlo simulations. These parameters were obtained from the full flare fits, where the decay of the QPPs was fitted with an exponential as this fit was far more stable than when a Gaussian decay was used (see Appendix \ref {section[histograms]}).

Figures \ref{figure[2dhist_p]} and \ref{figure[2dhist_phi]} show that $P_e$ and $\phi_e$ are anti-correlated with the flare amplitude $A_0$. In the full flare fits, shown in Figure \ref{figure[flare_split]} (with values given in Table \ref{table[total_fits_bands]}), the fitted value of $A_0$ is lower in the high-energy band than in the low-energy band, while both $P_e$ and $\phi_e$ are larger in the high-energy band than the low-energy band, which is consistent with the 2D histograms. This could indicate that the observed variation is indeed an artefact. However, we note that the observed parameter spaces are very different for the low- and high-energy bands and so therefore there is still a notable discrepancy. 

Furthermore, the 2D histograms show that, for the Monte Carlo simulations, $P_e$ and $\phi_e$ are positively correlated with $t_0$. However, although $P_e$ and $\phi_e$ are higher in the high-energy band than the low-energy band, $t_0$ is lower in low-energy band than the high-energy band. In other words, varying $t_0$ along the lines suggested by the correlation observed in the 2D histogram would not improve the agreement between the values of period and phase obtained for the two energy bands, again strengthening the assertion that the discrepancy is not an artefact of the analysis. The obtained period and phase show little correlation with $C$, which is substantially different in the two bands. 

\begin{figure*}
   \centering
    \includegraphics[width=0.7\textwidth]{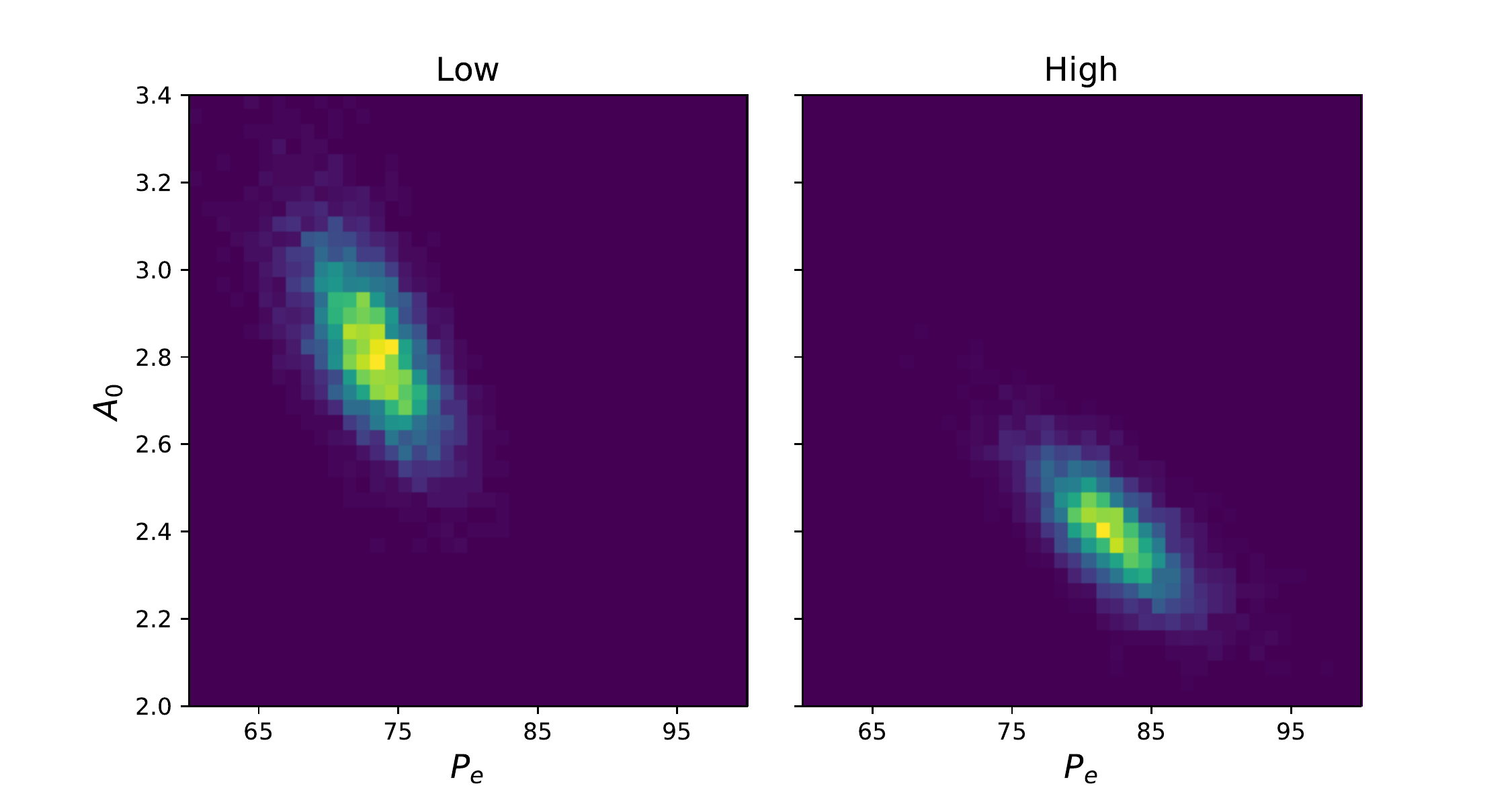}\\
	\includegraphics[width=0.7\textwidth]{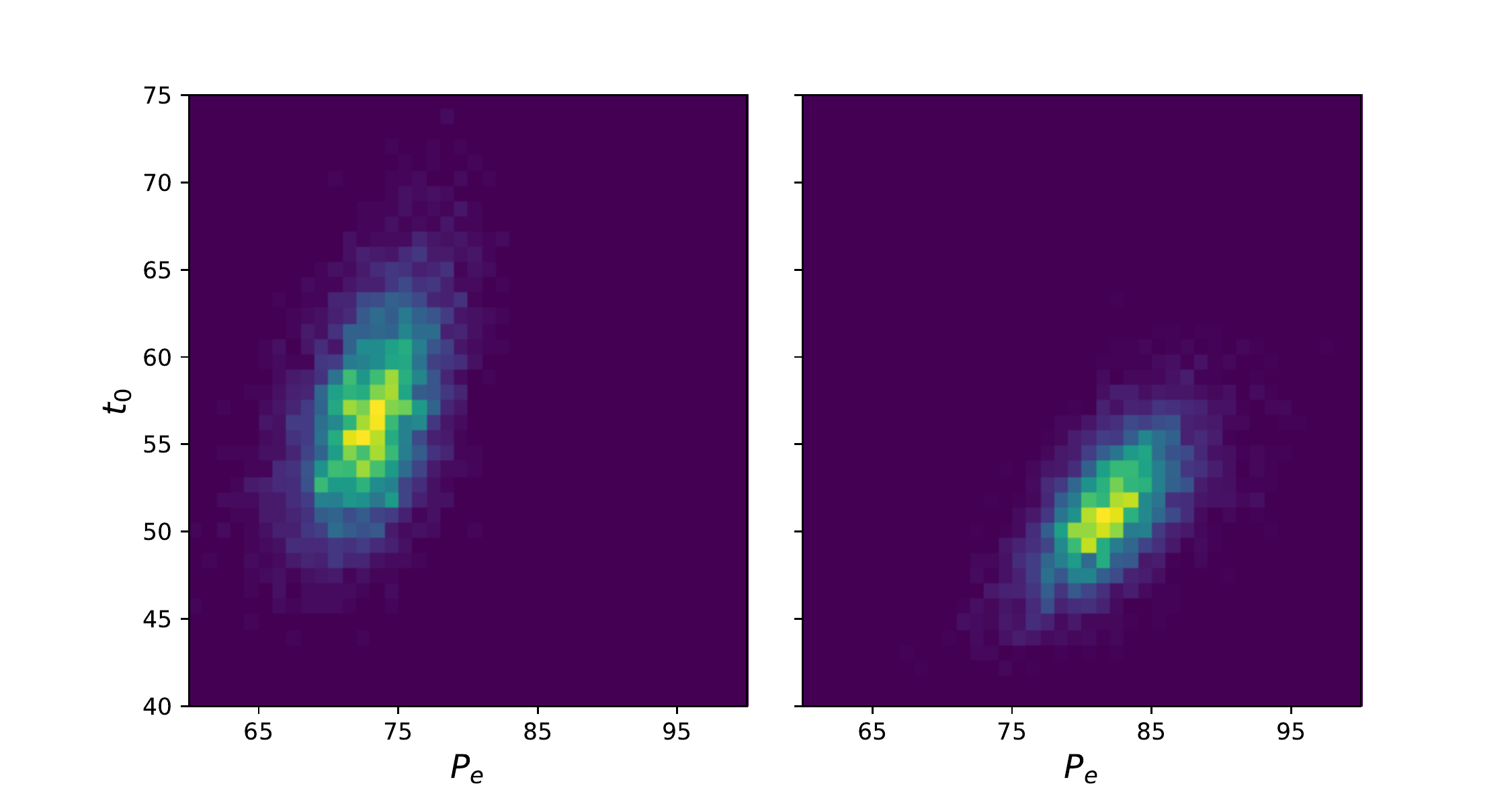}\\
	\includegraphics[width=0.7\textwidth]{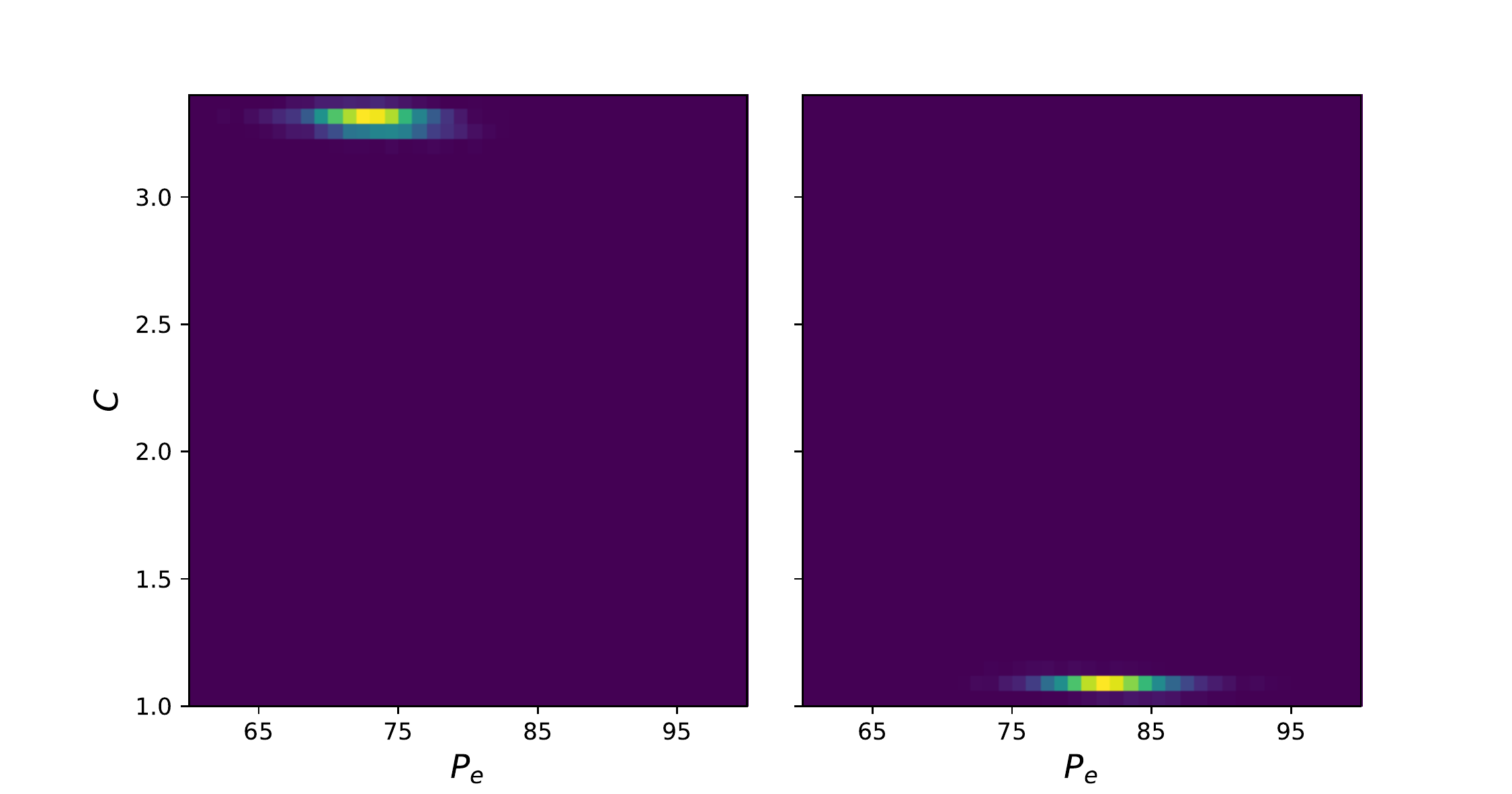}\\
   \caption{Two-dimensional histograms of full flare fit parameters obtained when QPP with exponentially-decaying amplitudes were fitted to data from congruent energy bands. Left-hand panels show the low-energy band (0.2-1.0\,keV) and right-hand panels show the high-energy band (1.0-12.0\,keV). In all panels, horizontal axes are fitted period of the QPP, $P_e$. In top row, vertical axes are amplitudes of the flare, $A_0$; in middle panels, vertical axes are decay times of the flare, $t_0$; and in bottom panels, vertical axes are constant offsets, $C$.}
              \label{figure[2dhist_p]}%
	\end{figure*}

\begin{figure*}
   \centering
    \includegraphics[width=0.7\textwidth]{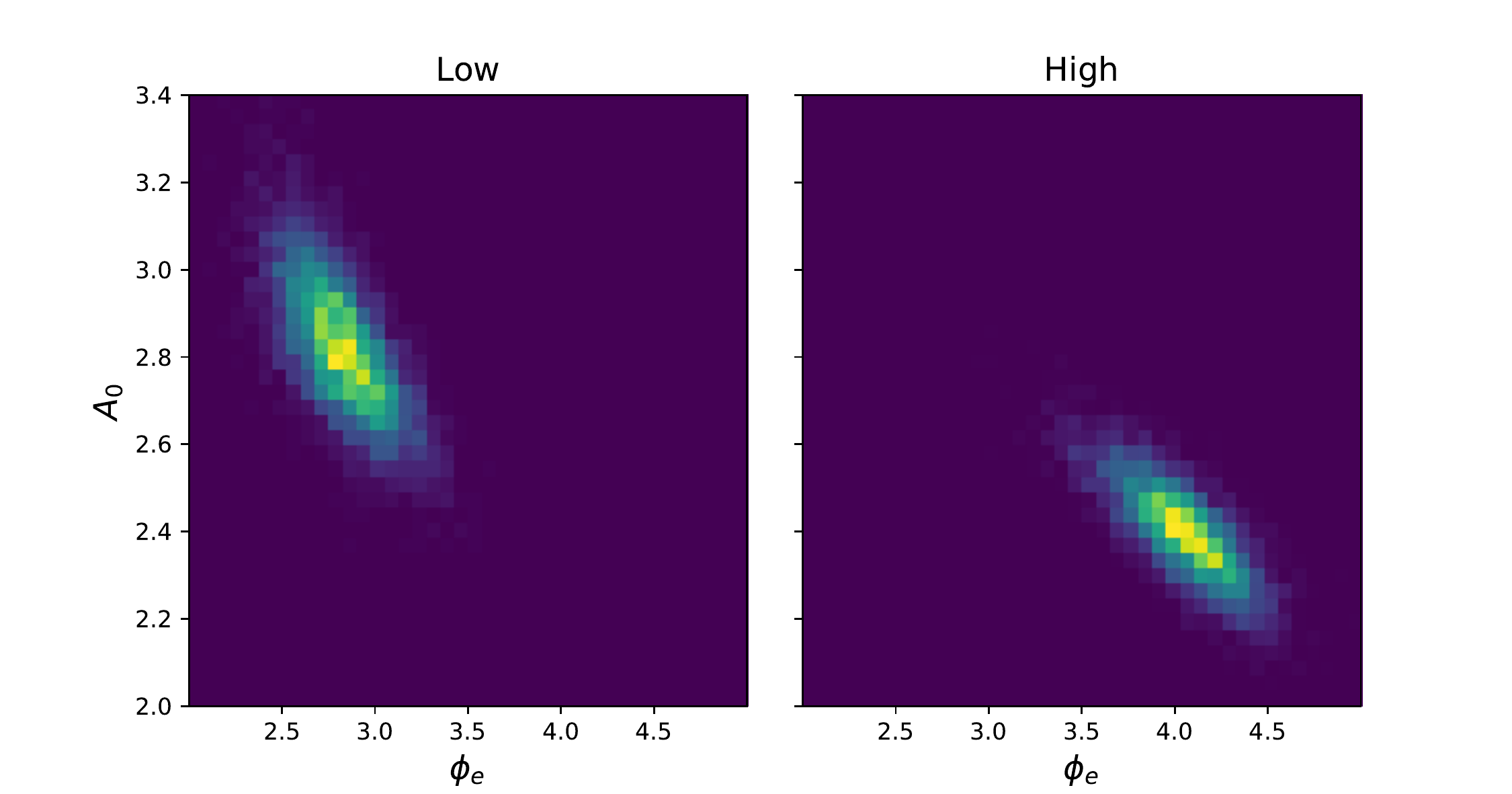}\\
	\includegraphics[width=0.7\textwidth]{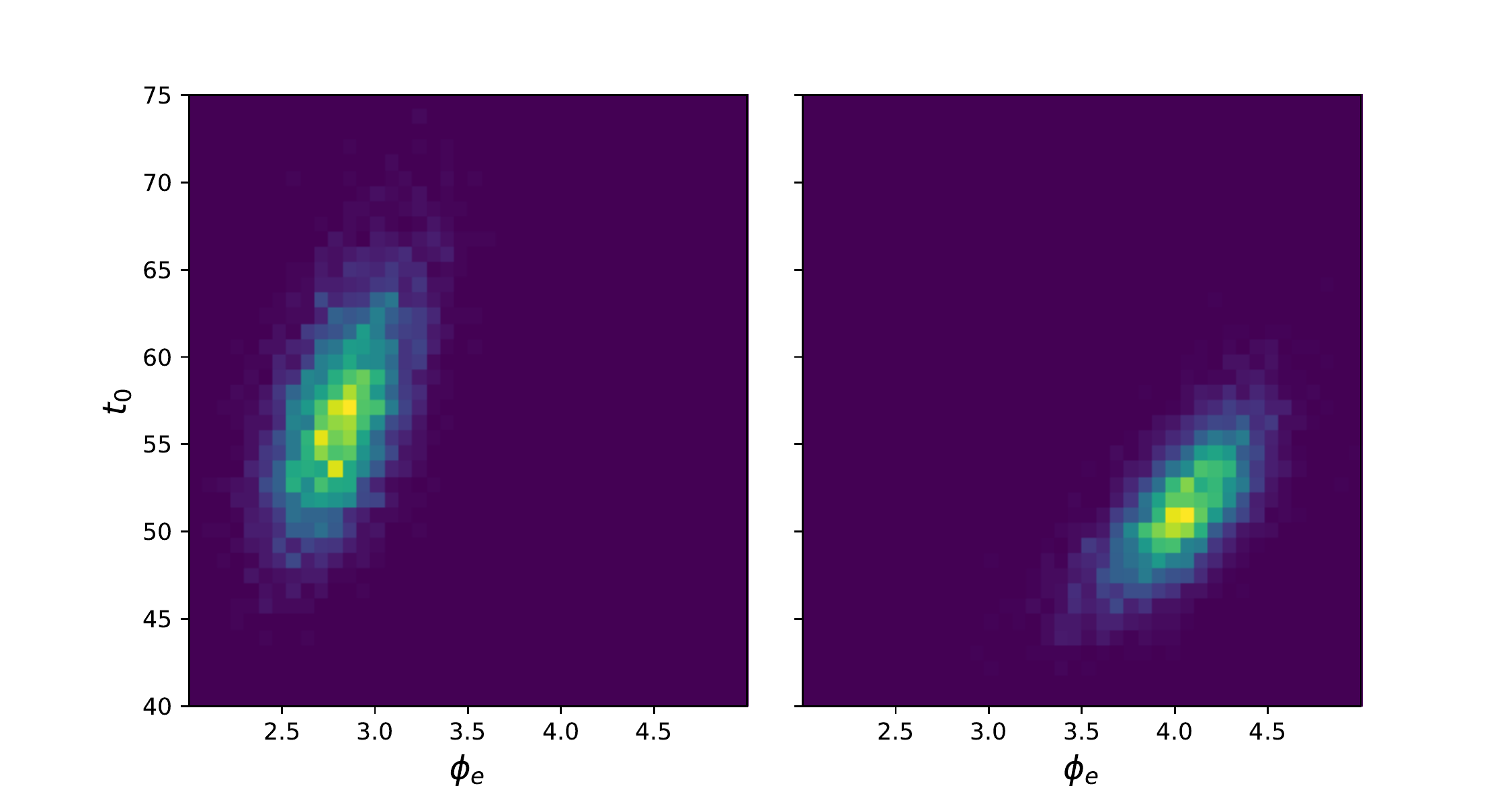}\\
	\includegraphics[width=0.7\textwidth]{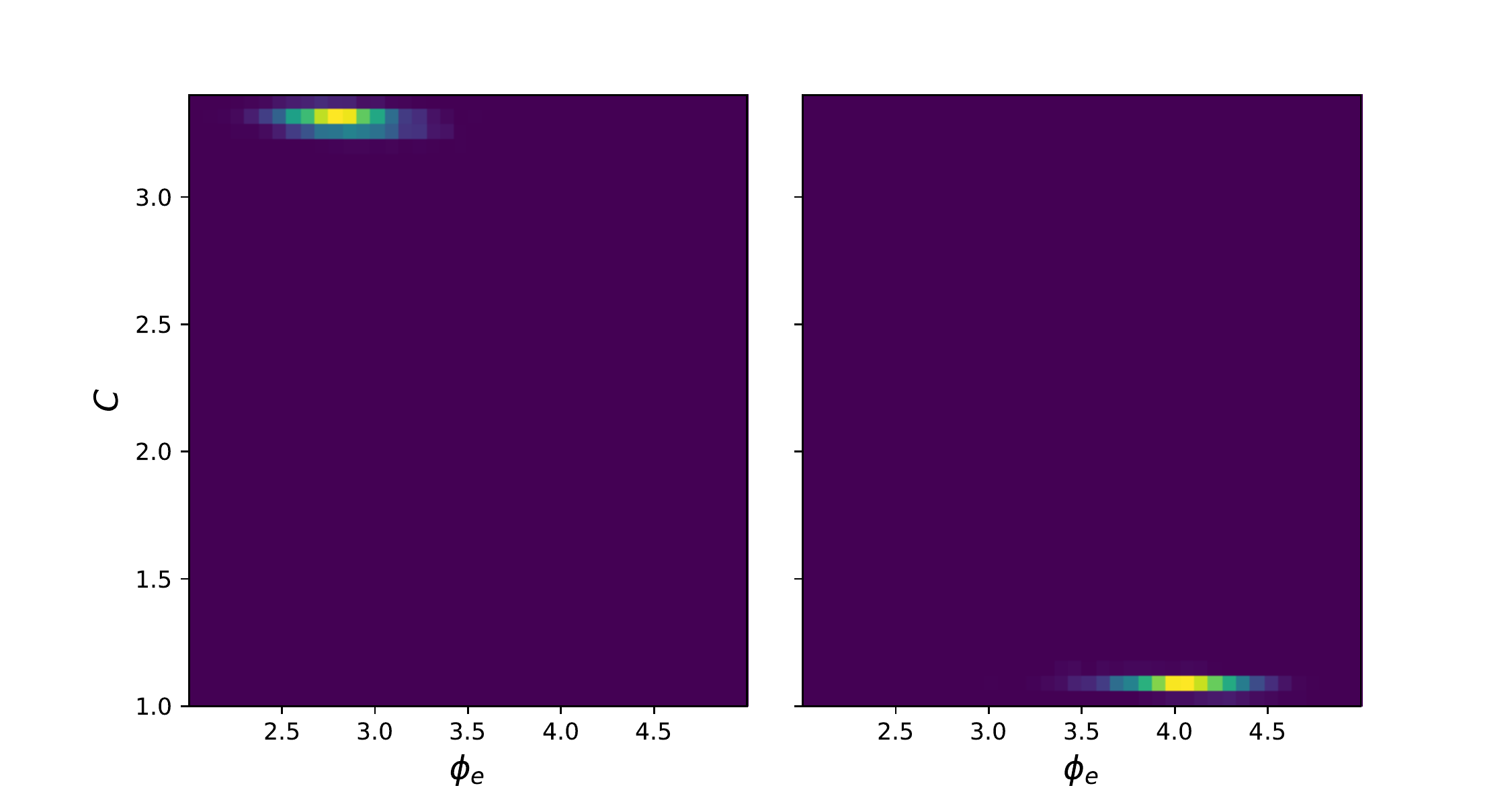}\\
   \caption{Two-dimensional histograms of full flare fit parameters obtained when QPP with exponentially-decaying amplitudes were fitted to data. As in Figure \ref{figure[2dhist_p]}, except that the horizontal axes are QPP phase, $\phi_e$.}
              \label{figure[2dhist_phi]}%
	\end{figure*}


\end{document}